\documentclass[dvipsnames]{article}

\pdfoutput=1

\usepackage[left=1.15in,right=1.15in,top=1in]{geometry}
        
\usepackage{graphicx,amsmath,amsthm,amssymb,mathtools,bbm}
\usepackage[outdir=./]{epstopdf}
\usepackage{indentfirst}
\usepackage{textcomp}
\DeclareMathOperator*{\argmin}{arg\,min}
\DeclareMathOperator*{\argmax}{arg\,max}
\newcommand{\bs}[1]{\boldsymbol{#1}}
\newcommand{\R}{\mathbbm{R}}
\newcommand{\pc}{PC}
\newcommand{\pce}{\pc{} expansion}
\newcommand{\uq}{UQ}
\newcommand{\sa}{SA}
\newcommand{\uqsa}{\uq{} and \sa{}}
\newcommand{\mc}{MC}
\newcommand{\cfp}{WAFP}
\newcommand{\odc}{ODC}
\newcommand{\potwo}{P\textsc{o}\textsubscript{2}}
\newcommand{\pfifty}{P\textsubscript{50}}

\newcommand{\otwo}{O\textsubscript{2}}
\newcommand{\pcotwo}{P\textsc{co}\textsubscript{2}}
\newcommand{\cotwo}{CO\textsubscript{2}}
\newcommand{\dpg}{DPG}
\newcommand{\twothreedpg}{2,3-DPG}
\newcommand{\ph}{pH}
\newcommand{\tmp}{$T$}
\newcommand{\shbotwo}{SHbO\textsubscript{2}}

\newcommand{\etal}{et al.}

\newcommand{\revision}[1]{{\leavevmode\color{black}{#1}}}
\newcommand{\revisionthree}[1]{{\leavevmode\color{BrickRed}{#1}}}
\renewcommand{\revisionthree}[1]{#1}

\usepackage{pgfplotstable}
\usepackage{booktabs}
\usepackage{array}
\usepackage{colortbl}
\usepackage[footnotesize,bf]{caption}
\usepackage{authblk}

\newcommand{\orgdiv}[1]{#1}
\newcommand{\orgname}[1]{#1}
\newcommand{\orgaddress}[1]{#1}
\newcommand{\country}[1]{#1}
\newcommand{\state}[1]{#1}

\title{Efficient sampling for polynomial chaos-based uncertainty quantification and sensitivity analysis using weighted approximate Fekete points\footnote{This work was supported in part by a Utah Space Grant Consortium Graduate Research Fellowship (KMB) and NIH award U24-EB029012.}}
\author[1,2]{Kyle M. Burk}
\author[3,4]{Akil Narayan}
\author[1,2]{Joseph A. Orr}

\affil[1]{\orgdiv{Department of Biomedical Engineering}, \orgname{University of Utah}, \orgaddress{\state{Utah}, \country{USA}}}
\affil[2]{\orgdiv{Department of Anesthesiology}, \orgname{University of Utah}, \orgaddress{\state{Utah}, \country{USA}}}
\affil[3]{\orgdiv{Mathematics Department}, \orgname{University of Utah}, \orgaddress{\state{Utah}, \country{USA}}}
\affil[4]{\orgdiv{Scientific Computing and Imaging Institute}, \orgname{University of Utah}, \orgaddress{\state{Utah}, \country{USA}}}

\begin{document}
\maketitle






\abstract{Performing uncertainty quantification (\uq{}) and sensitivity analysis (\sa{}) is vital when developing a patient-specific physiological model because it can quantify model output uncertainty and estimate the effect of each of the model's input parameters on the mathematical model.  By providing this information, \uqsa{} act as diagnostic tools to evaluate model fidelity and compare model characteristics with expert knowledge and real world observation.  Computational efficiency is an important part of \uqsa{} methods and thus optimization is an active area of research.  In this work, we investigate a new efficient sampling method for least-squares polynomial approximation, weighted approximate Fekete points (\cfp{}).  We analyze the performance of this method by demonstrating its utility in stochastic analysis of a cardiovascular model that estimates changes in oxyhemoglobin saturation response. Polynomial chaos (\pc{}) expansion using \cfp{} produced results similar to the more standard Monte Carlo in quantifying uncertainty and identifying the most influential model inputs (including input interactions) when modeling oxyhemoglobin saturation, \pce{} using \cfp{} was far more efficient.  These findings show the usefulness of using \cfp{} based \pce{} to quantify uncertainty and analyze sensitivity of a oxyhemoglobin dissociation response model.  Applying these techniques could help analyze the fidelity of other relevant models in preparation for clinical application.}


\maketitle



\section{Introduction}
\label{intro}
Performing uncertainty quantification (\uq{}) and sensitivity analysis (\sa{}) is vital when developing a patient-specific physiological model.  \uqsa{} can be used to quantify the relative effect of a model's input variables (individually or interactively) on the variability of the response of a physiological model.\cite{sudret2008global}  By doing so, \uqsa{} provide the basis for comparing model output behavior with expert knowledge and real world observations.  Results from this comparison can then be used to determine if the model is ready for clinical testing or if further adjustments are necessary.  

Although \uqsa{} were first used for assessing mathematical models of control and mechanical systems, they are also useful for assessing biological systems.\cite{geneser2008application} One method for ascertaining uncertainty and sensitivity with respect to varying input variables, \revision{when unable to compute the output distribution directly}, is a polynomial chaos (\pc{}) expansion.\cite{xiu2010numerical} A \pce{} \revision{represents} a model's output based on values of random inputs.\cite{crestaux2009polynomial} This prediction is stochastic and, compared with simpler approaches such as computationally demanding Monte Carlo (\mc{}) methods, \revision{can provide a more efficient representation of model output variability. A \pce{} can be more efficient because after a one-time construction cost is invested}, statistical quantities such as expected values, variances, higher-order moments, and variable sensitivities can all be computed via an inexpensive post-processing manipulation of the \pce{}.

In \pce{} methods, the PC surrogate is formed by seeding the surrogate construction algorithm with model evaluations. A PC surrogate is a multivariate polynomial function of a specified degree or order, with the model parameters serving as inputs to the polynomial. With respect to the number of degrees of freedom or expansion coefficients in the PC surrogate, it is recommended that a multiplicative sampling factor $f_S$ be used so that the minimum number of model evaluations is at least two times the number of expansion coefficients ($f_S = 2$).\cite{donders2015personalization}  Under this requirement, the computational expense grows with the number of model runs which in turn increases with the number of model parameters and the maximum polynomial order. Due to this limitation, once the number of model parameters or maximum polynomial order surpasses a certain threshold, \pce{} strategies may no longer be more efficient than \mc{} methods.  

The computational expense for constructing PC surrogates can be reduced by either reducing the number of model parameters, the polynomial order, or reducing the number of required model runs.  The number of model parameters can be reduced by first screening for important parameters before applying \pce{}.\cite{donders2015personalization}  This in effect reduces model dimensionality which in turn reduces the required number of model runs, \revision{although it should be noted that screening may miss important interactions and understimate variability}.  A different approach, for reducing the number of required model runs, is to use a sparse polynomial chaos expansion.\cite{blatman2010efficient}  This approach uses an iterative procedure that reduces the number of model runs by only adding basis terms that improve the \pce{}.\cite{blatman2010adaptive}  This in turn reduces the number of required model runs by reducing the number of unknown coefficients. One can investigate also decreases in the polynomial order of the PC model, but this effectively decreases the descriptive power of the surrogate, and a very small polynomial order may blunt the surrogate's expressivity \revision{and decrease the accuracy of quantities of interest}. The third strategy of reducing the number of model runs has received much attention in the literature, and is usually concerned with developing model query strategies that minimize the requisite number of model runs required to accurately construct a PC expansion. Our approach follows this third strategy.

Recently, Guo \textit{et al.} have introduced a new efficient sampling method for least-squares polynomial approximation, Christoffel-weighted approximate Fekete points (\cfp{}).\cite{guo_weighted_2018}  Rather than a multiplicative sampling factor this sampling method may only require an additive factor.  By using an additive factor, this approach may decrease computational expense by directly reducing the required number of model runs while still including all basis functions up to the maximum polynomial order specified. The method is flexible with respect to the number of model evaluations generated, the polynomial order, the number of model parameters, and the probability distributions associated to each parameter. In addition, the method requires only fairly simple linear algebra procedures to be implemented.

The purpose of this paper is to demonstrate the accuracy of a \cfp{} based \pce{} and its sensitivity indices when using an additive sampling factor. This is demonstrated by calculating the error of the \pce{} constructed using an additive sampling factor and by comparing its sensitivity indices with indices calculated using a multiplicative sampling factor as well as using \mc{} methods.  After demonstrating the accuracy of a \cfp{} based PC expansion, this paper then constructs and analyzes a \pce{} in order to analyze a model for describing changes in oxyhemoglobin dissociation response. The efficacy of the proposed approach on this example model suggests that it can be used to efficiently perform stochastic analysis for UQ on more general cardiovascular models.

The remainder of this paper is organized as follows.  The next section defines the notation that will be used throughout the paper.  Section \ref{sec:pce} explains the construction of a \pce{} and approximation of its coefficients.  In Section \ref{sec:wafp} the Christoffel-based weighted approximate Fekete point approach is presented.  Section \ref{sec:uqsa} presents how uncertainty and sensitivity characteristics were obtained from the \pce{}.  Sections \ref{sec:qa}-\ref{sec:cnvrg} detail the methods used for assessing a \pce{} and its indices. Section \ref{mc:methods} details the MC methods used for comparison while Section \ref{sec:model} describes the oxyhemoglobin dissociation response model.  The computational efficiency of \cfp{} based \pce{} is demonstrated in Section \ref{sec:results} by applying the methods to a oxyhemoglobin dissociation model.  Lastly, Section \ref{sec:discuss} concludes the paper by discussing the \cfp{} procedure and its application to the dissociation model.


\section{Notation}
\label{sa:notation}
Throughout this paper, an output of interest $Y$ for a model $f$ with $k$ inputs in $\mathbf{X}=(X_1, X_2,...,X_k)$ will be denoted as follows:  
\begin{equation}
Y=f(X_1,X_1,...,X_k)=f(\mathbf{X}),
\end{equation}
\noindent{}where $k$ is the number of independent stochastic input parameters $X_k$.  The output of interest $Y$ is the model's stochastic output and is uncertain because the inputs in $\mathbf{X}$ are uncertain themselves.  The stochastic inputs are contained by $\mathbf{X}$.  Here, we have defined the inputs, $X_1,X_2,...,X_k$, as uniform continuous variables which are mutually independent. The input parameter $X_i$ has probability density $\rho_i$. We assume that each input parameter $X_i$ has been linearly mapped to the domain $\Omega_i=[-1,1]$, so that each $X_i \in \Omega_i = [-1,1]$. For input parameters taking on bounded values, this is always possible.

When the model $Y$ is a physiological model, the domain $\Omega_i$ and the distribution of the input parameters will depend on the research question.  When exploring the effect of measurement uncertainty on the variance of a model output, the domain should be defined using a measure of variation such as standard deviation or 95\% confidence intervals.  If known, the distribution of the input parameters should match the known distribution of each type of measurement.  The effect of realistic physiological ranges on a model output can also be explored.  This analysis can help determine the expected variance in a model output across the populations of both healthy and unhealthy patients.  It can also determine the most influential input parameters on the model output.  For this type of analysis, the domain should be defined across the entire realistic physiological range and, if known, the input parameter distribution should match the distribution of experimental or clinical measurements.  If the distribution is unknown, the input parameters are usually assumed to have a uniform distribution.

Performing \pc{} assumes that $f(\bf{X})$ is smooth with respect to $\bf{X}$ across the sample space $\Omega_{\mathbf{X}} \coloneqq \times_{i=1}^k \Omega_i$. In other words, it assumes $f(\bf{X})$ has many derivatives with respect to $\bf{X}$.\cite{eck2016guide} Performing quality assessment of the \pce{} as detailed by Dubreuil \textit{et al.}\cite{dubreuil2014construction} and as implemented below can be used to verify the accuracy of the \pce{} approximation.  If quality assessment testing shows that the \pce{} approximation is accurate than this assumption can be considered valid. 

\revision{Performing a quality assessment for PC surrogates can be used to determine when the approach in this article is useful and when it is not.  If a PC emulator that accurately predicts the model can be constructed, then the procedures in this article will be beneficial for practitioners who wish to explore the model. However, some models are not smooth with respect to the input parameters $\bf{X}$, and thus building a straightforward PC surrogate will not be effective. Alternatives to PC emulators are Gaussian process emulators\cite{kennedy_bayesian_2001}, which may perform better in such situations.}

\section{Polynomial chaos expansion}\label{sec:pce}
\label{pce:methods}
Sensitivity measures for the model were estimated using the Python software toolkit \textit{UncertainSCI} \cite{uncertainsci} which implements a \cfp{}-design-based \pc{} method.  In this section we provide details of the methods \textit{pyopoly} uses for \pce{} construction and coefficient approximation.  Later, Section \ref{sec:qa} provides details of the methods \textit{pyopoly} uses to quantify uncertainty (expected value, variance, and prediction interval) and analyze sensitivity (main and total sensitivity indices).

\subsection{Polynomial chaos expansion construction}
A truncated \pce{} of the output $Y$ can be represented by the expression:
\begin{equation}\label{eq:pcexp}
  Y\approx \widetilde{f}(\mathbf{X})=\sum_{\alpha\in \mathcal{A}}c_{\alpha}\Phi_\alpha(\mathbf{X}),
\end{equation}
where $c$ are the expansion coefficients, $\Phi$ are the polynomials, $\alpha$ are multi-indices with nonzero entities, and $\mathcal{A}$ is the set of $N_p={k+p \choose k}$ multi-indices $\alpha$ for all polynomials with a maximal order $p$. \revisionthree{We have constructed $\{\Phi_\alpha\}_{\alpha \in \mathcal{A}}$ to be $L^2$ orthonormal polynomials with respect to density.}  When weighted by the joint density $\rho_{\bs{x}}$ of the random variable $\bs{X}$, i.e., they satisfy
  \begin{align*}
    \int_{\R^k} \Phi_\alpha(\bs{x}) \Phi_\beta(\bs{x}) \rho_{\bs{x}}(\bs{x}) \mathrm{d}\bs{x} = \delta_{\alpha,\beta},
  \end{align*}
  for all $\alpha, \beta \in \mathcal{A}$, where $\delta_{\alpha,\beta}$ is the Kronecker delta.

For our model $f$, we further assume a uniform distribution for the input distribution $\rho_i$ of the continuous random inputs $X_i$ for simplicity. Having assumed this distribution, we use a particular \pce{} basis, the Legendre polynomial family, for ease in computing and manipulating sensitivities.\cite{xiu2002wiener}  The construction of a \pce{} then proceeds by specifying a design of experiments over input variable space $\Omega_\mathbf{x}$.  The \pce 's maximal degree $p$ of the total degree polynomial approximation space is specified which results in a total of $N_p={k+p \choose k}$ polynomial basis elements. \revision{We note that our restriction to the uniform distribution in this paper does not limit the applicability of our PC emulator approach: all of our procedures generalize to non-uniform distributions.}

\subsection{Approximation of the expansion coefficients}
We construct the \pce{} with a weighted discrete least squares procedure.  Least squares approximation typically requires a multiplicative sampling factor $f_{S}^M$ to designate the number of sampling points where $N_S=f_{S}^MN_p$. Least squares requires a minimum $f_{S}^M=1$ sampling factor but generally $f_{S}^M \geq 2$ is recommended to obtain a good least squares minimization.\cite{eck2016guide,donders2015personalization} When using \cfp{}, we observe that least squares approximation may not require a multiplicative sampling factor but instead may be able to use a more lenient additive sampling factor $f_{S}^A$ where $N_S=N_p+f_{S}^A$. An additive sampling factor can be particularly efficient (i.e., requiring less data) when $p\gg 1$ or $k\gg 1$ since the number of polynomial basis elements grows algebraically with the maximal polynomial order and exponentially with the number of model parameters.\cite{blatman2009adaptive}  \revision{Using preliminary testing,} we have found that for \cfp{} based \pce{} methods, an additive sampling factor $f_{S}^A$ of 10 is typically sufficient for the model example we consider.  \revision{We note that our choice of $f_{S}^A = 10$ is made for the types of models we consider. While we have experienced that the robustness of this factor is fairly uniform for the simulations we have investigated, such a oversampling factors should be re-assessed for general situations.}

Explicitly, we assume that $N_S$ data points are available, $\left\{\bs{x}^{(i)}, y^{(i)} \right\}_{i=1}^{N_S}$, where $y^{(i)} \coloneqq Y(\bs{x}^{(i)})$. A weighted least squares procedure computes coefficients $c_\alpha$ in \eqref{eq:pcexp} satisfying a standard least squares problem,
\begin{align}\label{eq:wls-formulation}
  \{c_\alpha\}_{\alpha \in \mathcal{A}} = \argmin_{\{d_\alpha\}_{\alpha \in \mathcal{A}} \in \R^{N_p}} \sum_{i=1}^{N_S} w^{(i)} \left[ y^{(i)} - \sum_{\alpha \in \mathcal{A}} c_\alpha \Phi_\alpha(\bs{x}^{(i)}) \right]^2,
\end{align}
where the weights $w^{(i)}$ and the sample grid locations $\bs{x}^{(i)}$ must be specified; the problem above is a standard least squares objective function, and can be solved explicitly via standard linear algebra techniques. The special distinction of the \cfp{} strategy is the choice of the weights and grid locations. We describe this choice next.

\section{Weighted approximate Fekete points}\label{sec:wafp}

Once \eqref{eq:wls-formulation} is solved, the \pce{} surrogate $f$ in \eqref{eq:pcexp} is determined and may be used for predictive and analysis purposes. The method of \cfp{} determines a choice of the weights $w^{(i)}$ and sample locations $\bs{x}^{(i)}$ in \eqref{eq:wls-formulation} \cite{guo_weighted_2018}. We briefly describe this choice here. The weights are uniquely defined by the sample locations, and are chosen as
\begin{align*}
  w^{(i)} = \left[ \sum_{\alpha \in \mathcal{A}} \Phi_\alpha(\bs{x}^{(i)})^2 \right]^{-1}.
\end{align*}
Thus, once an experimental design for the locations $\bs{x}^{(i)}$ is determined, then the weights are specified. The first step in determining the sample locations is construction of a large ``training" grid, 
\begin{align}\label{eq:training-grid}
  \left\{ \bs{z}^{(1)}, \ldots, \bs{z}^{(M)} \right\} &\subset \R^{k},  M \gg N_S.
\end{align}
We will choose $M \sim K N_p$ for some constant $K$, but at present we simply assume that this grid is present and discuss in Section \ref{ssec:training-grid} how this grid is generated.

In what follows we detail how the \cfp{} design points are chosen. The procedure is completed in two steps: in the first step we choose $N_p$ sample points via one scheme, and in the second step the reamining $N_S - N_p$ points are chosen via a different procedure. These steps are described in brief in the following sections. The first step is identical to the original \cfp{} procedure \cite{guo_weighted_2018}. The second step is a modification of the original approach for oversampling, and is more computationally efficient than the original approach. Although several other methods for experimental design exist, the \cfp{} strategy is simple and consistently produces stable \pce{} approximations in practice\cite{guo_weighted_2018}.

\subsection{Selection of the first $N_p$ points}
The true model $Y$ is \textit{not} evaluated on the training grid. Instead, the following model-independent matrix is formed,
\begin{align}\label{eq:B-def}
  \bs{B} = \left( \begin{array}{cccc} \psi_{\alpha_1}\left( \bs{z}^{(1)}\right) & \psi_{\alpha_1}\left( \bs{z}^{(2)}\right) & \cdots & \psi_{\alpha_1}\left( \bs{z}^{(M)} \right) \\
                                      \psi_{\alpha_2}\left( \bs{z}^{(1)}\right) & \psi_{\alpha_2}\left( \bs{z}^{(2)}\right) & \cdots & \psi_{\alpha_2}\left( \bs{z}^{(M)} \right) \\
                                                \vdots & \vdots & \ddots & \vdots \\
  \psi_{\alpha_{N_p}}\left( \bs{z}^{(1)}\right) & \psi_{\alpha_{N_p}}\left( \bs{z}^{(2)}\right) & \cdots & \psi_{\alpha_{N_p}}\left( \bs{z}^{(M)} \right)
\end{array}\right) \in \R^{N_p \times M},
\end{align}
where $(\alpha_1, \ldots, \alpha_{N_p})$ is any enumeration of the elements in $\mathcal{A}$. The functions $\psi_{\alpha}$ are defined as
\begin{align*}
  \psi_\alpha\left(\bs{z}^{(i)}\right) = \sqrt{w^{(i)}} \Phi_\alpha \left(\bs{z}^{(i)}\right).
\end{align*}
The matrix $\bs{B}$ is used to choose the sample locations $\{\bs{x}^{(i)}\}_{i=1}^{N_S}$. Precisely, a column-pivoted $Q R$ decomposition is performed on $\bs{B}$, and the first $N_p$ column indices determined by this procedure identify a size-$N_p$ subset of $\{\bs{z}^{(i)}\}_{i=1}^M$ that defines the first $N_p$ sample locations. The selection is based on analysis ensuring robust mathematical properties \cite{bos_computing_2010,cohen_optimal_2017}. Note that a column-pivoted $Q R$ decomposition simply performs column selection via greedy determinant maximization, so that in the language of statistical design of experiments, this is a type of approximate $D$-optimal design. Note, however, that it is not a standard $D$-optimal design since a \textit{weighted} determinant is maximzed.

\subsection{Choosing the remaining points}
The remaining $N_S - N_p$ samples are chosen via a slightly different procedure using a greedy determinantal maximization, which is slightly different than the oversampling procedure described in \cite{guo_weighted_2018}. The previous section detailed the choosing of $\{\bs{x}^{(i)}\}_{i=1}^{N_p}$ via a $Q R$ decomposition that is effectively greedy determinant maximization. In this section we describe the choice of the set $\{\bs{x}^{(i)}\}_{i=N_p+1}^{N_S}$, which is still greedy determinant maximization, but must be implemented slightly differently.

The explicit iterative choice of $\bs{x}^{(i)}$ for $i > N_p$ is given by 
\begin{align}\label{eq:x-determinant-max}
  \bs{x}^{(i)} &= \argmax_{\bs{z} \in \{ \bs{z}^{(m)}\}_{m=1}^M \backslash \{ \bs{x}^{(m)}\}_{m=1}^{N_p}} \det(\bs{G}^{(i)}(\bs{z})), & i &= N_p+1, \ldots, N_p + N_S,
\end{align}
where $\bs{G}^{(i)}(\bs{z})$ and $\bs{B}^{(i)}$ are, respectively, $i \times i$ and $N_p \times i$ matrices defined by
\begin{align*}
  \bs{G}^{(i)}(\bs{z}) &\coloneqq \bs{B}^{(i)}(\bs{z}) \left(\bs{B}^{(i)}(\bs{z})\right)^T, &
  \bs{B}^{(i)}(\bs{z}) &\coloneqq \left( \begin{array}{cccc} \psi_{\alpha_1}\left( \bs{x}^{(1)}\right) & \cdots & \psi_{\alpha_1}\left( \bs{x}^{(i-1)}\right) & \psi_{\alpha_1}\left( \bs{z} \right) \\
  \psi_{\alpha_2}\left( \bs{x}^{(1)}\right) & \cdots & \psi_{\alpha_2}\left( \bs{x}^{(i-1)}\right) & \psi_{\alpha_2}\left( \bs{z} \right) \\
                                                \vdots & \ddots & \vdots & \vdots \\
  \psi_{\alpha_{N_p}}\left( \bs{x}^{(1)}\right) & \cdots & \psi_{\alpha_{N_p}}\left( \bs{x}^{(i-1)}\right) & \psi_{\alpha_{N_p}}\left( \bs{z} \right)
\end{array}\right)
\end{align*}
Note that the explicit determinant above can be optimized more easily than the formulas suggest. If we define the $N_p$-vector $\bs{\psi}(\bs{z})$ and the $N_p \times N_p$ matrix $\bs{\Psi}(\bs{z})$,
\begin{align*}
  \bs{\psi}(\bs{z}) &\coloneqq \left( \begin{array}{c} \psi_{\alpha_1}(\bs{z}) \\ \psi_{\alpha_2}(\bs{z}) \\ \vdots \\ \psi_{\alpha_{N_p}}(\bs{z}) \end{array} \right), & 
  \bs{\Psi}(\bs{z}) &\coloneqq \bs{\psi}(\bs{z}) \bs{\psi}(\bs{z})^T,
\end{align*}
then we see that
\begin{align*}
  \bs{G}^{(i)} &= \bs{\Psi}(\bs{z}) + \bs{G}_i, & \bs{G}_i &\coloneqq \sum_{r=1}^{i-1} \bs{\Psi}(\bs{x}^{(r)}),
\end{align*}
where $\bs{G}_i$ does not depend on $\bs{z}$. Thus, by the matrix determinant lemma,
\begin{subequations}\label{eq:quick-determinant}
\begin{align}
  \det(\bs{G}^{(i)}(\bs{z}^{(j)})) = \det(\bs{G}_i) \left[ 1 + \bs{\psi}\left(\bs{z}^{(j)}\right)^T \bs{G}_i^{-1} \bs{\psi}\left(\bs{z}^{(j)}\right) \right].
\end{align}
Therefore, maximizing the determinant in \eqref{eq:x-determinant-max} is equivalent to maximizing the bracketed term above, which involves only a single matrix-vector and a single vector-vector multiplication, assuming $\bs{G}_i^{-1}$ is stored from the previous iteration. In addition, once $\bs{x}^{(i)}$ is computed from \eqref{eq:x-determinant-max}, then $\bs{G}^{-1}_{i+1}$ can be computed from $\bs{G}_i^{-1}$ without matrix inverses via the Sherman-Morrison formula:
\begin{align}
  \bs{G}_{i+1}^{-1} = \bs{G}_i^{-1} - \frac{\bs{G}_i^{-1} \bs{\psi}\left(\bs{x}^{(i)}\right) \bs{\psi}\left(\bs{x}^{(i)}\right)^T \bs{G}_i^{-1}}{1 + \bs{\psi}\left(\bs{x}^{(i)}\right)^T \bs{G}_i^{-1} \bs{\psi}\left(\bs{x}^{(i)}\right)}
\end{align}
\end{subequations}
Therefore, the maximization \eqref{eq:x-determinant-max} can be performed by simply using the update formulas \eqref{eq:quick-determinant} in an iterative fashion. This completes our description of how the \cfp{} points $\{\bs{x}^{(i)} \}_{i=1}^{N_S}$ are chosen; our remaining task is to describe how the training grid is chosen.

\subsection{Choosing the training grid}\label{ssec:training-grid}
The quality of the \cfp{} nodes described in the previous subsections depends on the choice of the training grid in \eqref{eq:training-grid}. The \cfp{} procedures are determinant maximizing procedures that use linear algebraic operations. These operations in turn are most accurate when the matrix $\bs{B}$ is well-conditioned. Thus, we seek to generate a good set of candidate samples so that the condition number of $\bs{B}$ is as small as possible. 

In principle this can accomplished by a space-filling design, but this requires exploration of the entire parameter domain, which can be prohibitively expensive when $k$ is large. To circumvent this, we leverage a least squares result regarding random sampling. Suppose that $\{\bs{z}^{(i)}\}_{i=1}^M$ are independent and identically distibuted random samples from the probability density
\begin{align}\label{eq:induced-distribution}
  \rho_{\mathcal{A}}(\bs{x}) \coloneqq \frac{1}{N_p} \sum_{\alpha \in \mathcal{A}} \Phi_\alpha(\bs{x})^2 \rho_{\bs{x}}(\bs{x})
\end{align}
The density $\rho_{\mathcal{A}}$ is called the induced distribution\cite{narayan_computation_2018}, and is a property only of the joint distribution $\rho(\bs{x})$ and the polynomial index set $\mathcal{A}$.  With the $\bs{z}^{(i)}$ distributed according to $\rho_{\mathcal{A}}$, then taking a training set of size $M$ satisfying for any $C > 0$ ensures that $\bs{B}$ is well-conditioned with high probability \cite{cohen_optimal_2017}. More precisely, the condition number inequality $\kappa\left(\bs{B} \bs{B}^T\right) \leq 3$ holds with probability exceeding $1 - 2 M^{-1-C}$. Thus, an essential multiplicative behavior of $M \sim K N_p$ for some constant $K$ ensures that $\bs{B}$ is well-conditioned. This is the procedure we use to choose our training grid. Note that efficient algorithms to generate random samples from a large class of induced distributions are available\cite{narayan_computation_2018}, but even a simple rejection sampling algorithm can easily generate samples from $\rho_{\mathcal{A}}$ using samples of $\rho_{\bs{x}}$.

\begin{figure}[htbp]
	\centering
	\includegraphics[width=0.7\textwidth{}]{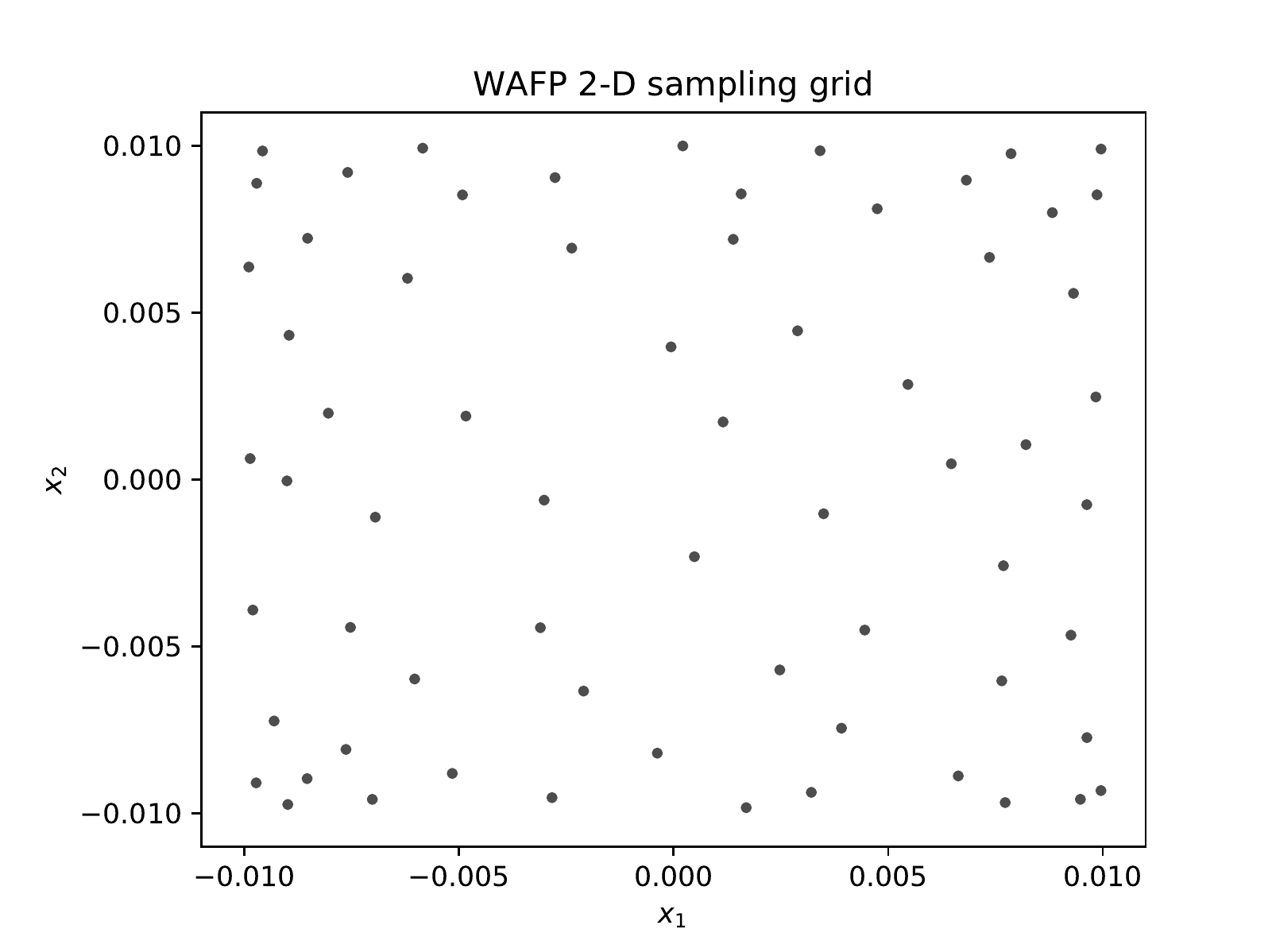}
	\caption{Two-dimensional grid generated using the weighted approximate Fekete points procedure.}
	\label{fig:wafp}
\end{figure}

For our \cfp{} training sets, we choose our candidate sets as random samples from $\rho_{\mathcal{A}}$ but note that alternative methods also generate good candidate sets\cite{guo_weighted_2018}.  \revisionthree{Sampling points when using the \cfp{} procedure with maximal polynomial order $p=10$ and sampling factor $f_S^A=0$ for the Oakley and O'Hagen\cite{oakley2002bayesian} two-dimensional function are shown in Figure \ref{fig:wafp}.}


\subsection{Alternatives to WAFP least squares}

In the sections above we have chosen the WAFP least squares strategy to compute the coefficients $c_\alpha$ in the PC expansion \eqref{eq:pcexp}. This is actually two choices: that of \text{least squares} approximation, and of the \revision{WAFP} experimental design. We provide brief motivation of these choices in this section, along with discussion of alternative formulations that one can consider when constructing PC expansions. \revision{We do not give a comprehensive review of sampling strategies; more strategies and more complete discussions are available in the literature\cite{narayan_stochastic_2015,cohen_optimal_2017}.}

The least squares strategy, i.e., the formulation \eqref{eq:wls-formulation} defining the coefficients $c_\alpha$, could be replaced by an interpolation strategy (if $N_S = N_p$), or by more complicated methods such as sparse approximation with compressive sampling \cite{candes_stable_2006,rauhut_mathematical_2013}. Our choice of the least squares formulation is largely motivated by the fact that simple, efficient, and stable algorithms can solve this problem, whereas more complicated algorithms can be required for alternative formulations. In addition, least squares formulations are intimately connected to statistical regression models, which have a long history of theoretical and algorithmic development \cite{draper_applied_1998}.

Having selected least squares as the approximation technique, we can now discuss the particular sample design, i.e., the selection of the grid $\left\{\bs{x}^{(i)}\right\}_{i=1}^{N_S}$. In this article we cannot do justice to the voluminous historical work on sampling and design of experiments for polynomial approximation; instead we focus on a discussion of recent advances and approaches, and seek mainly to appropriately align and compare our approach with alternative existing methods. For least squares approximations, we discuss below Monte Carlo methods, potential-theory-based least squares designs, and designs obtained via optimization.

\subsubsection{(Quasi-) random sampling designs for regression-PC}
Perhaps the simplest approach for designing a sampling scheme is with randomization, which produces a random sample configuration. This approach is flexible, simple, and, generalizable. In this paper we have assumed that $\bs{X}$ is a random vector comprised of mutually independent, uniformly distributed random variables. Monte Carlo\revisionthree{-type} methods and algorithms apply to this case, but also apply straightforwardly without assumptions of independence or uniform distributions. For \revisionthree{random sample design PC} methods, one constructs samples $\left\{\bs{x}^{(i)}\right\}_{i=1}^{N_S}$ as independent and identically distributed (iid) realizations of the random variable $\bs{X}$. One then computes the PC coefficients from \eqref{eq:wls-formulation} with $w^{(i)} = 1$ for all $i$. The analysis in \cite{cohen_stability_2013} reveals that, if $N_S$ is large enough, then \eqref{eq:wls-formulation} is a \revision{well-conditioned regression design with high probability and computes a PC expansion of near-optimal accuracy, i.e., computes a PC expansion whose $L^2$ error is almost as small as the best possible PC expansion, where ``best" is measured in the $L^2$ sense.} The drawback is that for most distributions of $\bs{X}$, $N_S$ must be rather large compared to $N_P$. For example, in our case of independent uniform random variables, we require $N_S \gtrsim N_P^2$ \revisionthree{using random, independent samples}. \revision{Note that this behavior of $N_S$ relative to $N_P$ guarantees convergence with high probability, but often in practice one observes that $N_S = f_S^M N_P$ for some constant $f_S^M$ can empirically yield convergence.}

A slight variant of the \revisionthree{random sampling} algorithm, using importance sampling to bias the sampling distribution, can substantially improve convergence results. Such a weighted \revisionthree{randomized sampling} procedure requires only that $N_S\log(N_S) \gtrsim N_P$ in order to obtain \revision{linear algebraic stability of the weighted least-squares problem \eqref{eq:wls-formulation} and convergence of $\tilde{f}$ to the best polynomial approximation \cite{cohen_optimal_2017}}. The optimal biased distribution from which to choose samples is the induced distribution shown in \eqref{eq:induced-distribution}. \revision{The analysis then suggests that $N_S \gtrsim f_S^M N_P$ samples are required (ignoring logarthmic factors) to obtain accurate approximations \cite{cohen_optimal_2017}; in contrast the WAFP approach we employ empirically requires only $N_S = N_P + f_S^A$ samples \cite{guo_weighted_2018}. However, we do not yet have rigorous theory that mathematically demonstrates the superiority of WAFP designs compared to randomized designs. We justify our choice of WAFP (in contrast to iid realizations), along with our recommended choice of $f_S^A = 10$ for the model considered in this manuscript in Figure \ref{fig:l2error_AM}.
}

\subsubsection{Designs based on pluripotential theory}
The second class of least squares designs that have become popular are those based on potential-theory concepts. (In the multivariate case these are pluripotential theory concepts.) In the framework of least squares approximations, large polynomial degree asymptotics for the distribution of an optimal sampling design are known\cite{bloom_asymptotics_2011}. More precisely, for a fixed polynomial degree $k$, one can define a sampling design as optimal if it maximizes a particular weighted matrix determinant; the large-$k$ limit of such designs must distribute according to a weighted pluripotential equilibrium measure. Similar notions of such large-degree convergence to equilibrium measures are known in more abstract settings \cite{bloom_convergence_2010}. This theory has inspired computational methods for least squares that exploit such a connection to pluripotential theory\cite{narayan_christoffel_2017}. These pluripotential-theory-based approaches can effectively generate good designs for large polynomial degree approximation \textit{if} explicit forms for weighted pluripotential equilibrium measures is known. The disadvantages of these approaches are that (i) the precise form of such weighted equilibrium measures are not explicitly known even in relatively simple cases, making the development of computational methods difficult in general, and (ii) the optimality of such designs is true in the large degree limit; it is not clear how effective these approaches are for small-to-medium polynomial degrees. In contrast, the WAFP algorithm we explore generates empirically good designs for an arbitrary polynomial subpsace, and does so without any knowledge of equilibrium measures. However, the theory for WAFP designs is still nascent and lacks some of the asymptotically-optimal properties of some other least squares designs.

\subsubsection{Generation of designs via optimization}
The last class of least squares designs we discuss are those generated by numerical optimization. Our WAFP approach fits best in this category. Such approaches attempt to compute a design by numerically optimizing a quality metric of the design. Such metrics can be moment matching conditions \cite{mehrotra_generating_2013,keshavarzzadeh_numerical_2018} or matrix determinants \cite{shin_nonadaptive_2016}, but a wide variety of such metrics can be devised \cite{narayan_constructing_2013,hadigol_least_2018}. More recently, compression techniques based on algorithms inspired by Carath\'eodory's Theorem and Tchakaloff's theorem have been used to reduce a very large discrete set of samples into a smaller set of samples over which least squares procedures are stable \cite{8024337,box2019near,bos_generating_2018}. These procedures aim to generate parsimonious least squares designs that matches moments of a large discrete sample set instead of a continous density. In contrast our WAFP approach matches the moments of the continuous density. In addition, the WAFP approach requires only standard linear algebra procedures along with the explicit formulas in Section \ref{ssec:training-grid}, whereas many of the previously-mentioned strategies require non-trivial optimization routines for implementation.

\section{Uncertainty quantification and sensitivity analysis}\label{sec:uqsa}
Uncertainty characteristics of expected value and variance are quantified directly using the \pce{}.  The prediction interval is estimated by approximating the probability density function for $Y$ ($\rho_{Y}$) using the \pce{} surrogate model and \mc{} methods.

The expected value and variance of $\rho_{Y}$ are obtained using the \pce{}.  The expected value is the \pce's first expansion coefficient:
\begin{equation}
\mathbb{E}[Y]\approx{}\int_{_\Omega{}_x}\sum_{\alpha\in\mathcal{A}}c_\alpha\Phi_\alpha(\mathbf{x})\rho_\mathbf{x}(\mathbf{x})d\mathbf{x}=c_0.
\end{equation}
The variance is the sum of squared expansion coefficients minus the first expansion coefficient:
\begin{equation}
  \mathbb{V}[Y]\approx{}\mathbb{E}[(\widetilde{f}(\mathbf{X})-\mathbb{E}[Y])^2]=\int_{_\Omega{}_\mathbf{x}}(f(\mathbf{x})-c_0)^2\rho_\mathbf{x}(\mathbf{x})d\mathbf{x}=\int_{_\Omega{}_x}\Bigg(\sum_{\alpha\in\mathcal{A}}c_\alpha\Phi_\alpha(\mathbf{x})\Bigg)^2\rho_\mathbf{x}(\mathbf{x})d\mathbf{x}-c_0^2.
\end{equation}
Since percentiles and prediction intervals cannot be derived directly from the \pce{} of $Y$, we apply the \mc{} method to find them.  A 95\% prediction interval is created from the output realizations when evaluating the \pce{} for $10,000N_s$ new input samples.

We use Sobol indices to determine which parameters have little or no effect on the output $Y$ and which parameters heavily influence $Y$.\cite{sobol2001global2001} Both main (first-order) and total indices are obtained using the \pce{}.  Doing so requires approximating the total variance in $Y$ by
\begin{equation}
\mathbb{V}[Y]\approx{}\sum_{\alpha\in\mathcal{A}}\mathbb{V}[c_\alpha\Phi_\alpha(\mathbf{X})].
\end{equation}
Main sensitivity ($S_i$) is a unitless quantity between 0 and 1 that measures the relative variance a certain variable (without interaction) contributes to the total output variance $\mathbb{V}[Y]$. For example, $S_1$ measures only the variance that $X_1$ imparts alone, and does not measure the variance contributed by, e.g., the pair ($X_1$, $X_2$) or any other pairs or higher-order tuples where $X_1$ is involved.  $S_1$ could represent the expected reduction in $\mathbb{V}[Y]$ that would be achieved if $X_1$ were set to its standard physiological value.\cite{eck2016guide}  The main sensitivity index ($S_i$) is calculated as:
\begin{align}
S_i\approx \frac{1}{\mathbb{V}[Y]}\sum_{\alpha\in\mathcal{A}_i}\mathbb{V}[c_\alpha\Phi_\alpha],\hskip 5pt \textrm{where} \hskip 5pt \mathcal{A}_i=\{\alpha|\alpha_i>0\wedge\alpha_j=0 \hskip 10pt \forall j\neq i\}.
\end{align}
Second-order sensitivity $S_{ij}$ is also a unitless quantity between 0 and 1 that measures the relative variance that the interaction of two certain variables contributes to the total output variance $\mathbb{V}[Y]$. For example, $S_{1,2}$ measures the variance contributed by the pair ($X_1$, $X_2$). Higher order indices also exist, and these indices measure the variance contributed by higher-order tuples.\cite{eck2016guide}

Total sensitivity ($S_{i}^T$) measures the total relative variance a variable subset contributes, even if other variables participate in this contribution. $S_{1}^T$ sums the first-order effect that $X_1$ imparts alone ($S_1$), along with the higher-order effects by, e.g., $S_{1,2}$ and any other pairs or higher-order tuples involving $X_1$.  If there is any variance imparted by the interaction between parameters, the sum of $S_{i}^T$ for all parameters will be greater than one. The total sensitivity index can be approximated by:
\begin{align}
S_{i}^T\approx \frac{1}{\mathbb{V}[Y]}\sum_{\alpha\in\mathcal{A}_{i}^T}\mathbb{V}[c_\alpha\Phi_\alpha],\hskip 5pt \textrm{where} \hskip 5pt \mathcal{A}_{i}^T=\{\alpha|\alpha_i>0\}.
\end{align}
$S_i$ is useful for ranking uncertain inputs and determining which inputs contribute to model output variability.  $S_{i}^T$ is useful for analyzing the interaction between uncertain inputs and determining which input interactions contribute to model output variability.  If $S_i$ and $S_{i}^T-S_i$ are near zero this indicates that a particular input and its interactions do no contribute to model output variability.  The input can then be set to a standard physiological value without affecting model fidelity.

\section{Quality assessment}\label{sec:qa}

In order to properly evaluate \cfp-based \pce{}, the \pce{} in (\ref{eq:pcexp}), as well as the main and total sensitivity indices obtained from it, should be validated \revision{using a specific model (see Section 9)}.  For this evaluation, an error measure for the \pce{} as well as the sensitivity indices should be calculated.  \revision{This can be done by comparing the results for the \pce{} and sensitivity indices to reference values.  For the analysis in this paper reference values were obtained using a WAFP-based \pce{} with a $p=10$ maximal order \pce{} and $f_{S,M} = 2$ sampling factor.} This assessment is performed in accordance with the previous work of Dubreuil \textit{et al.} and Donders \textit{et al.} which provide further detail in their manuscripts.\cite{donders2015personalization,dubreuil2014construction}

\subsection{Quality of the polynomial chaos expansion}
The error in the \pce{} (descriptive error) can be quantified using $R^2$, the coefficient of determination

\begin{align}\label{eq:deserr}
  \epsilon_{R^2}=1-R^2=\frac{\sum_{i=1}^{N_s}\Big(y^{(i)}-\widetilde{f}_{P_{50}}(\mathbf{x}^{(i)})\Big)^2}{\sum_{i=1}^{N_s}\big(y^{(i)}-\overline{y}\big)^2},
\end{align}

\noindent{}where $\overline{y}_j$ is the sample mean of the realizations of $Y_j$.  The descriptive error $\epsilon_{R^2}$ represents the residual variance as a fraction of the total variance.  

The \pce{} can be validated using leave-one-out cross-validation, \revision{which computes the validation error using the same $N_s$ data points that are collected for construction of the \pce{}}.  The error in the \pce 's predictions (predictive error) can be quantified using $Q^2$, the validation coefficient

\begin{align}\label{eq:prederr}
  \epsilon_{Q^2}=1-Q^2=\frac{\sum_{i=1}^{N_s}\Big(y^{(i)}-\widetilde{f}_{P_{50}}^{(-i)}(\mathbf{x}^{(i)})\Big)^2}{\sum_{i=1}^{N_s}\big(y^{(i)}-\overline{y}\big)^2},
\end{align}

\noindent{}\revision{where $\widetilde{f}_{P_{50}}^{(-i)}$ represents a \pce{} constructed with all but the $i^{th}$ sample (i.e., with $N_s - 1$ samples). Thus, $N_s$ \pce{} expansions are constructed to compute this metric.} The predictive error $\epsilon_{Q^2}$ represents the predicted variance as a fraction of the total variance. 

\subsection{Quality of the sensitivity indices}
The relative error of the estimated (main and total) sensitivity indices $S$ with respect to the reference sensitivity indices $\hat{S}$ can be used to quantify the accuracy of the sensitivity indices.  The relative error is

\begin{align}\label{eq:relerr}
\epsilon_\delta=\frac{|S_i-\hat{S_i}|}{\hat{S_i}},
\end{align}

\noindent{}where the index $i=\{1,2,...,2k\}$, because both main and total indices are considered.  For the analysis in this paper reference values $\hat{S}$ are obtained using a WAFP-based \pce{} with a $p=10$ maximal order \pce{} and $f_{S,M} = 2$ sampling factor.

\section{Assessment of convergence}\label{sec:cnvrg}
\label{sec:cnvg}
To assess convergence, we set the multiplicative sampling factor $f_{S}^M$ to 2 and the additive factor $f_{S}^A$ to 10, \revision{both of which were WAFP \pce s}.  Descriptive error ($\epsilon_{R^2}$) and predictive error ($\epsilon_{Q^2}$) can then be calculated for $2N$ \pce s ($N$ multiplicative and $N$ additive) generated with maximal \pce{} orders $p=\{1,2,3,...,N\}$.  Descriptive error is calculated using (\ref{eq:deserr}) and predictive error using (\ref{eq:prederr}).

\pce s of a high maximal order $p=N$ or less can be constructed to estimate the convergence of error and variation in the sensitivity indices. Sensitivity indices are obtained from each of the $2N$ \pce s and are then evaluated for accuracy and precision compared to the reference analysis ($p=N+1$ and $f_{S}^M=2$).  Accuracy of the sensitivity indices is evaluated by calculating relative error using (\ref{eq:relerr}).

\section{Monte Carlo method}
\label{mc:methods}

For \uq{}, the expected value, variance, and prediction interval are estimated from the evaluations of $Y$ using standard methods.  For \sa{} we used Saltelli's alternative \mc{} procedure as implemented in the R software toolbox \textit{sensitivity} \cite{iooss2018sensitivity}.  This method used two sampling matrices of $N$ samples each drawn independently of each other.  For a four-dimensional model, $k=4$ matrices are generated from the two sampling matrices for a total of six matrices.  Model evaluations are then calculated for each row in all six matrices.  Further detail is provided in the literature for the interested reader.\cite{saltelli2002making,eck2016guide}

\subsection{Assessment of convergence}

We used Saltelli's \mc{} method with $N=100,000$ samples to quantify uncertainty and analyze sensitivity for comparison.  To evaluate the convergence of Saltelli's \mc{} method, the relative error of the sensitivity indices was also calculated using (\ref{eq:relerr}) for $N=\{10,100,1000,10000\}$.  \revision{Reference values were obtained using a WAFP-based \pce{} with a $p=10$ maximal order \pce{} and $f_{S,M} = 2$ sampling factor.}

\section{Oxyhemoglobin dissociation model}
\label{sec:model}

\subsection{Background}
\begin{figure}[htbp]
\centering
\includegraphics[width=0.8\textwidth{}]{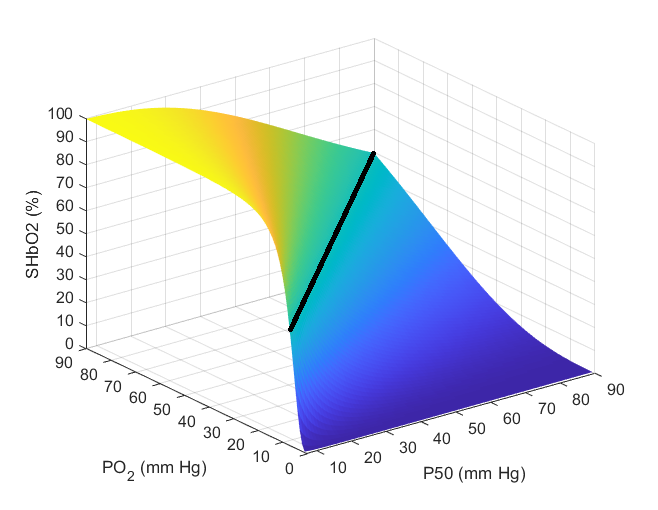}
\caption{Changes in oxyhemoglobin dissociation curve shape and position. The black solid line indicates how the partial pressure at which 50\% of hemoglobin are saturated with oxygen (\pfifty{}) changes.  \pfifty{}, and thus the oxyhemoglobin dissociation curve, varies from patient to patient where oxygen affinity and \pfifty{} have an inverse relationship (oxygen infinity increases as \pfifty{} decreases and vice versa).}
\label{fig:odc_standard}
\end{figure}

The oxyhemoglobin dissociation curve (\odc{}) describes the relationship between the partial pressure of oxygen in the blood (\potwo{}) and the percent of hemoglobin saturated with oxygen (\shbotwo{}).  This relationship varies from patient to patient.\cite{hemmings2006foundations} The position not the shape of the \odc{} varies, where for a given \potwo{} hemoglobin increases oxygen uptake as the curve shifts to the left and decreases uptake as the curve shifts to the right.\cite{hemmings2006foundations,severinghaus1958oxyhemoglobin,kelman1966digital} \pfifty{}, the \potwo{} at which hemoglobin is 50\% saturated (Figure \ref{fig:odc_standard}), is the inflection point of the \odc{} and thus is commonly used to represent the position of the entire \odc{}.  Determining a  patient-specific \pfifty{} is helpful as it can help predict the \shbotwo{} at which the patient's oxyhemoglobin saturation will transition from declining slowly to rapidly.

\pfifty{} is determined by four factors (three chemical and one physical).\cite{winslow1983simulation}  These four factors are the concentration of the hydrogen ion (\ph{}), temperature (\tmp{}), the partial pressure of carbon dioxide (\pcotwo{}), and the concentration of 2,3-diphosphoglycerate (\dpg{}).\cite{oudemans1996analysis} These factors influence \pfifty{} either directly or indirectly through each other.  Many groups have studied these factors to create models capturing each parameter's influence on \pfifty{}.\cite{kelman1966digital,winslow1983simulation,severinghaus1979simple,siggaard1984mathematical,buerk1986simplified,matejak2015adair} These groups have assessed the effect of these parameters as well as their synergistic effects using experimental data and computational methods in order to model transport and exchange of oxygen in the lungs and tissue.\cite{dash2010erratum} Most recently, Dash, Korman, and Bassingthwaighte described and evaluated an equation of state which incorporates the influence of all four variables (both individually and in combination) on \pfifty{}.\cite{dash2016simple}

A \uqsa{} of Dash's equation of state, which has 4 input parameters and 4 polynomial equations, could formally quantify the uncertainty of \pfifty{} and its sensitivity to variations in \ph{}, \tmp{}, \pcotwo{}, and \twothreedpg{}, thus providing a means for ranking these variables and their interactions from most to least significant.

One utility of \uqsa{} is that they can be used to analyze the fidelity of a model by comparing model characteristics to expert knowledge and real world observation.  Using \uqsa{} for this purpose can help determine if a model is prepared for clinical evaluation.  If the model is not ready, \uqsa{} can be used to determine which characteristics of the model's output do not align with theory and can provide insight and guidance for aligning model characteristics with theory.

The oxyhemoglobin dissociation model is suitable for highlighting this particular utility of \uqsa{}.  Oxyhemoglobin dissociation was first described in the early 1900s and has been discussed extensively in the literature.  The primary physical and chemical factors effecting oxyhemoglobin dissociation are well-established.  The influence of these factors on oxyhemoglobin dissociation has been studied both experimentally in the laboratory as well through observing pathophysiology of adjustments to different climates, situations, and diseases.  This broad scope of knowledge on oxyhemoglobin dissociation makes the oxyhemoglobin dissociation model an ideal candidate for demonstrating the benefits of applying \uqsa{} to evaluate model fidelity in preparation for transitioning to clinical use.

\subsection{Calculating \pfifty{}}

Dash \textit{et al.} analyzed Buerk and Bridges' model and experimental data to determine curve-fit polynomials for characterizing the relationship between \pfifty{} and changes in \ph{}, \tmp{}, \pcotwo{}, and \twothreedpg{} \revision{[concentration]}.\cite{dash2016simple,dash2010erratum}  They determined these equations by varying one variable while the other three were fixed at standard values.  Their results showed that the best-fit polynomial for the relationship between \pfifty{} and changes in each variable is given by:
\begin{align}\label{eq:P50-dash-model}
P_{50,\Delta{}\mathbf{Q_i}} = P_{50}^S + \sum_{j=1}^3 \mathbf{Q}_{ij} (\mathbf{Q}_i - Q^S)^j,
\end{align}
where the vector $\mathbf{Q} = \{pH, T, PCO_2, [DPG]\}$ contains the physiological levels of each of the four variables.  $\mathbf{Q}_{ij}$ represents the coefficients for each of these four variables.  These coefficients are given in Table \ref{tab:dash-coeffs}. The quantities with superscript ``S'' denote the standard physiological value for that variable (see Table \ref{table:standard}).  Previous studies \cite{buerk1986simplified,dash2016simple,kelman1966digital,severinghaus1979simple,siggaard1984mathematical} have shown when combining simultaneous changes in more than one variable, \pfifty{} is given by:
\begin{equation}\label{eq:P50-combined}
P_{50} = P_{50}^S\prod_{i=1}^{4}\Big(\frac{P_{50,\Delta \mathbf{Q}_i}}{P_{50}^S}\Big)
\end{equation}
Combining (\ref{eq:P50-dash-model}) and (\ref{eq:P50-combined}) gives the overall description for how \pfifty{} changes in response to varying physiological conditions:
\begin{equation}\label{eq:P50prod}
P_{50} = P_{50}^S\prod_{i=1}^{4}\Big(\frac{P_{50}^S+\sum_{j=1}^{3}\mathbf{Q}_{ij}(\mathbf{Q}_i-Q^S)^j}{P_{50}^S}\Big).
\end{equation}
\pfifty{} is the output of interest for the oxyhemoglobin dissociation model.  It is uncertain due to interpatient differences in \ph{}, \tmp{}, \pcotwo{}, and \twothreedpg{} which are uncertain themselves \revision{due to measurement uncertainty and stochastic variation over time and space within individuals.  The analysis for this portion of the experiment focuses  on the uncertainty and variability across the entire population.}  When subjected to uncertainty, we can express the four-parameter mathematical model ($f_{P_{50}}$) described by (\ref{eq:P50prod}) in black-box form as:
\begin{equation}
P_{50}=f_{P_{50}}(pH,T,PCO_2,\revision{[}DPG\revision{]})=f_{P_{50}}(\mathbf{X}),
\end{equation}
\noindent{}
where $P_{50}$ is the stochastic output representing \pfifty{} which in this case is a scalar value.  $P_{50}$ is the random field of interest and thus is a function of $\bf{X}$ which we have denoted by $f_{P_{50}}(\bf{X})$.  The stochastic inputs are contained by $\mathbf{X}=[pH,T,PCO_2,\revision{[}DPG\revision{]}]$.

We have defined the physiological sample space ($\Omega_i$) for each input in Table \ref{table:range}.  These are the realistic physiological ranges for each parameter.  By analyzing these ranges, we have considered the case where each parameter's uncertainty is across the entire realistic physiological range.  Therefore, the physiological range was the 100\% confidence interval for that given input.  This in effect evaluates the uncertainty of the output due to the inherent physiological variability of each variable in both healthy and unhealthy patient populations.

\subsection{Polynomial Chaos Expansion}

We analyzed the role of \ph{}, \tmp{}, \pcotwo{}, and \twothreedpg{} on the uncertainty of \pfifty{} by constructing a \cfp{} based \pce{}.  (\ref{eq:P50prod}) provides a model which maps the four input variables \ph{}, \pcotwo{}, \tmp{}, and \twothreedpg{} \revision{[concentration]} to the value of \pfifty{}.  Constructing a \pce{} provides a surrogate model from which uncertainty and sensitivity can be extracted directly.

We defined the inputs, $pH,T,PCO_2,\revision{[}DPG\revision{]}$, as uniform continuous variables which are mutually independent.  We used an additive sampling factor $f_{S}^A = 10$ and maximal polynomial order $p=3$ which we confirmed were acceptable with the pre-processing analysis performed in Sections \ref{sec:qa}-\ref{sec:cnvg} and in Appendix A.  This resulted in a \pce{} of $N_p={4+3 \choose 4}=35$ basis elements constructed from $N_s=35+10=45$ model runs.  The model's 95\% prediction intervals were then estimated by evaluating the \pce{} using $10,000N_s=450,000$ new input samples \revision{of each} input parameter.  For comparison, we calculated model output uncertainty and sensitivity indices using \mc{} methods as detailed in Section \ref{mc:methods}.

\subsection{Global \uqsa{}}
\label{methods:globuqsa}
We evaluated the output of the oxyhemoglobin dissociation model to evaluate the utility of using \pce{} for this model, as opposed to \mc{} methods.  We varied the value for each parameter, one at a time, within the physiologically realistic parameter space.  We did this both while holding the other input parameters to their standard values and while allowing the other input parameters to vary randomly within their sample space.  



\pgfplotstableset{
  columns/j/.style={int detect, column type=r, column name=$j$},
  columns/ph/.style={sci, sci zerofill, sci sep align, precision=3, sci 10e, column name=$pH_j$},
  columns/pco/.style={sci, sci zerofill, sci sep align, precision=3, sci 10e, column name=$PCO2_j$},
  columns/dpg/.style={sci, sci zerofill, sci sep align, precision=3, sci 10e, column name=$[DPG]_j$},
  columns/t/.style={sci, sci zerofill, sci sep align, precision=3, sci 10e, column name=$T_j$},
  every head row/.style={before row=\toprule,after row=\midrule},
  every last row/.style={after row=\bottomrule}
}
\begin{table}[htbp]
  \begin{center}
  {\scriptsize
\caption{Value of coefficients $Q_j$ in the $P_{50}$ model (Eq. \ref{eq:P50-dash-model}) from Dash \etal{}\cite{dash2016simple}}\label{tab:dash-coeffs}
\centering
\pgfplotstabletypeset{
 j  ph    pco     dpg     t
 1  -25.535 1.273e-1 795.63  1.435
 2      10.646  1.083e-4        19660.89        4.163e-2
 3      -1.764  0       0       6.86e-4
}
  }
  \end{center}
\end{table}

\begin{table}[htbp]
\caption{Standard values of the oxyhemoglobin dissociation model}
\label{table:standard}
\centering
  {\scriptsize
\begin{tabular}{llll}
\toprule
Parameter & Description & Value & Unit  \\
\midrule
pH$_S$ & Standard pH in plasma & 7.4 & Unitless\\
\tmp{}$_S$ & Standard temperature of blood & 37 & \textsuperscript{o}C \\
\pcotwo{}$_{,S}$ & Standard partial pressure of \cotwo{} in blood & 40 & mmHg\\
\dpg{}$_S$ & Standard 2,3-DPG concentration in RBCs & 4.65e\textsuperscript{-3} & M \\
\pfifty$_{,S}$ & \pfifty{} at pH$_S$, \tmp{}$_S$, \pcotwo{}$_S$, and \dpg{}$_S$ & 26.8 & mm Hg\\
\bottomrule
\end{tabular}
  }
\end{table}

\begin{table}[htbp]
\caption{Uncertainty ranges ($\Omega_i$) tested for the oxyhemoglobin dissociation model}
\label{table:range}
\centering
  {\scriptsize
\begin{tabular}{lllccl}
\toprule
Nr. & Symbol & Uncertainty Range & Unit  \\
\midrule
1 & pH & 6.0-8.2 & Unitless\\
2 & T & 20-43 & \textsuperscript{o}C \\
3 & \pcotwo{} & 10-100 & mm Hg\\
4 & \dpg{} & 1e\textsuperscript{-3}-10e\textsuperscript{-3} & M \\
\bottomrule
\end{tabular}
  }
\end{table}

\begin{table}[htbp]
\caption{Results of the UQ of \pfifty{} for \mc{} and \pc{} with uncertainty ranges across the entire physiological range.}
\label{table:uq}
\centering
  {\scriptsize
\begin{tabular}{lccc}
\toprule
Method & Mean & Standard deviation & 95\% Prediction interval \\
 & [mmHg] & [mmHg] & [mmHg]\\
\midrule
MC & 33.92 & 23.35 & [7.33, 94.73]\\
PC & 34.26 & 23.15 & [7.58, 94.16]\\
\bottomrule
\end{tabular}
  }
\end{table}


\begin{table}[htbp]
	\caption{Polynomial model sensitivities}
	\label{tab:polymdl}
	\centering
        {\scriptsize
	\begin{tabular}{lccccc}
		\toprule
		Sensitivity & Analytical & \multicolumn{4}{l}{Order of PC expansion}\\
		index & value & \cline{1-4} \\
		& & $p=3$ & $p=4$ & $p=5$ & $p=6$\\
		\midrule
		$S_1$ & 0.2747 & 0.2765 & 0.2744 & 0.2748 & 0.2747 \\
		$S_2$ & 0.2747 & 0.2741 & 0.2761 & 0.2748 & 0.2747 \\
		$S_3$ & 0.2747 & 0.2747 & 0.2745 & 0.2748 & 0.2747 \\
		$S_{12}$ & 0.0549 & 0.0533 & 0.0556 & 0.0549 & 0.0549 \\
		$S_{13}$ & 0.0549 & 0.0537 & 0.0545 & 0.0547 & 0.0549 \\
		$S_{23}$ & 0.0549 & 0.0571 & 0.0547 & 0.0550 & 0.0549 \\
		$S_{123}$ & 0.0110 & 0.0106 & 0.0100 & 0.0110 & 0.0110 \\
		&&&&&\\
		$N_p$ & & 20 & 35 & 56 & 84\\
		$N_s$ & & 30 & 45 & 66 & 94\\
		\bottomrule
	\end{tabular}
        }
\end{table}

\begin{table}[htbp]
	\caption{Ishigami function sensitivities}
	\label{tab:ish}
	\centering
        {\scriptsize
	\begin{tabular}{llllll}
		\toprule
		Index & Analytical & \multicolumn{4}{l}{Order of PC expansion}\\
		& value & \cline{1-4} \\
		& & $p=3$ & $p=5$ & $p=7$ & $p=9$\\
		\midrule
		$S_1$ & 0.3138 & 0.3558 & 0.3289 & 0.3161 & 0.3166 \\
		$S_2$ & 0.4424 & 0.0020 & 0.2839 & 0.4151 & 0.4386 \\
		$S_3$ & 0      & 0.1054 & 0.0072 & 0.0016 & 0.0000 \\
		$S_{12}$ & 0   & 0.0686 & 0.0465 & 0.0054 & 0.0001 \\
		$S_{13}$ & 0.2436 & 0.4543 & 0.2295 & 0.2315 & 0.2441 \\
		$S_{23}$ & 0   & 0.0052 & 0.0565 & 0.0120 & 0.0002 \\
		$S_{123}$ & 0  & 0.0085 & 0.4749 & 0.0184 & 0.0003 \\
		&&&&&\\
		$S_{1}^T$ & 0.5574 & 0.8873 & 0.6524 & 0.5713 & 0.5611 \\
		$S_{2}^T$ & 0.4424 & 0.0845 & 0.4344 & 0.4509 & 0.4392 \\
		$S_{3}^T$ & 0.2436 & 0.5735 & 0.3407 & 0.2635 & 0.2447 \\
		&&&&&\\
		\multicolumn{2}{l}{$N_p$} & 20 & 56 & 120 & 220\\
		\multicolumn{2}{l}{$N_s$} & 30 & 66 & 130 & 230\\
		\bottomrule
	\end{tabular}
        }
\end{table}

\begin{table}[htbp]
	\caption{Full G-function sensitivities}
	\label{tab:sobolfull}
	\centering
        {\scriptsize 
	\begin{tabular}{lcc}
		\toprule
		Index & Analytical & PC-based\\
		& solution & solution ($p=2$) \\
		\midrule
		$S_1$ & 0.6037 & 0.5134 \\
		$S_2$ & 0.2683 & 0.2299 \\
		$S_3$ & 0.0671 & 0.0311 \\
		$S_4$ & 0.0200 & 0.0412 \\
		$S_5$ & 0.0055 & 0.0046 \\
		$S_6$ & 0.0009 & 0.0098 \\
		$S_7$ & 0.0002 & 0.0194 \\
		$S_8$ & 0.0000 & 0.0034 \\
		&&\\
		$S_{1}^T$ & 0.6342 & 0.5402 \\
		$S_{2}^T$ & 0.2945 & 0.2782 \\
		$S_{3}^T$ & 0.0756 & 0.0765 \\
		$S_{4}^T$ & 0.0227 & 0.0789 \\
		$S_{5}^T$ & 0.0062 & 0.0104 \\
		$S_{6}^T$ & 0.0011 & 0.0447 \\
		$S_{7}^T$ & 0.0003 & 0.0538 \\
		$S_{8}^T$ & 0.0000 & 0.0645 \\
		\bottomrule
	\end{tabular}
        }
\end{table}

\begin{table}[htbp]
	\caption{Reduced G-function sensitivities}
	\label{tab:sobolred}
	\centering
        {\scriptsize
	\begin{tabular}{lclll}
		\toprule
		Index & Full model & \multicolumn{3}{l}{Reduced Model}\\
		& (eight parameters) & & & \\
		& \cline{1-4}\\
		& Analytical & $p=3$ & $p=5$ & $p=7$ \\
		\midrule
		$S_1$ & 0.6037 & 0.6216 & 0.5623 & 0.5696 \\
		$S_2$ & 0.2683 & 0.1887 & 0.2590 & 0.2403 \\
		$S_3$ & 0.0671 & 0.0714 & 0.0708 & 0.0760 \\
		$S_4$ & 0.0200 & 0.0172 & 0.0220 & 0.0106 \\
		&&&&\\
		$S_{12}$ & 0.0224 & 0.0252 & 0.0230 & 0.0202 \\
		$S_{13}$ & 0.0056 & 0.0235 & 0.0125 & 0.0127 \\
		$S_{14}$ & 0.0017 & 0.0146 & 0.0042 & 0.0072 \\
		$S_{23}$ & 0.0004 & 0.0102 & 0.0058 & 0.0090 \\
		$S_{24}$ & 0.0005 & 0.0048 & 0.0090 & 0.0068 \\
		$S_{34}$ & 0.0005 & 0.0078 & 0.0148 & 0.0051 \\
		&&&&\\
		$S_{1}^T$ & 0.6342 & 0.6900 & 0.6171 & 0.6438 \\
		$S_{2}^T$ & 0.2945 & 0.2431 & 0.3070 & 0.3096 \\
		$S_{3}^T$ & 0.0756 & 0.1238 & 0.1168 & 0.1385 \\
		$S_{4}^T$ & 0.0227 & 0.0593 & 0.0648 & 0.0624 \\
		&&&&\\
		\multicolumn{2}{l}{$N_p$} & 35 & 126 & 330 \\
		\multicolumn{2}{l}{$N_s$} & 45 & 136 & 340 \\
		\bottomrule
	\end{tabular}
        }
\end{table}


\section{Results}\label{sec:results}
\label{results}

\subsection{\pce{} convergence and quality}

After constructing \pce s for $p=\{1,2,3,...,\revision{11}\}$, we assessed the quality of the \pce s using a validation test set. \revision{The validation test set is generated independently of the experimental design and WAFP training set.}  \revisionthree{We recall that from Section \ref{ssec:training-grid} that the WAFP training set is generated from the density $\rho_{\mathcal{A}}$ in \eqref{eq:induced-distribution}. The validation set is generated as independent random samples from the density $\rho$.} The discrepancy on this validation set was computed for the \pce{} prediction of \pfifty{} compared with the true value. 

\begin{figure}[htbp]
\centering
\includegraphics[width=0.49\textwidth{}]{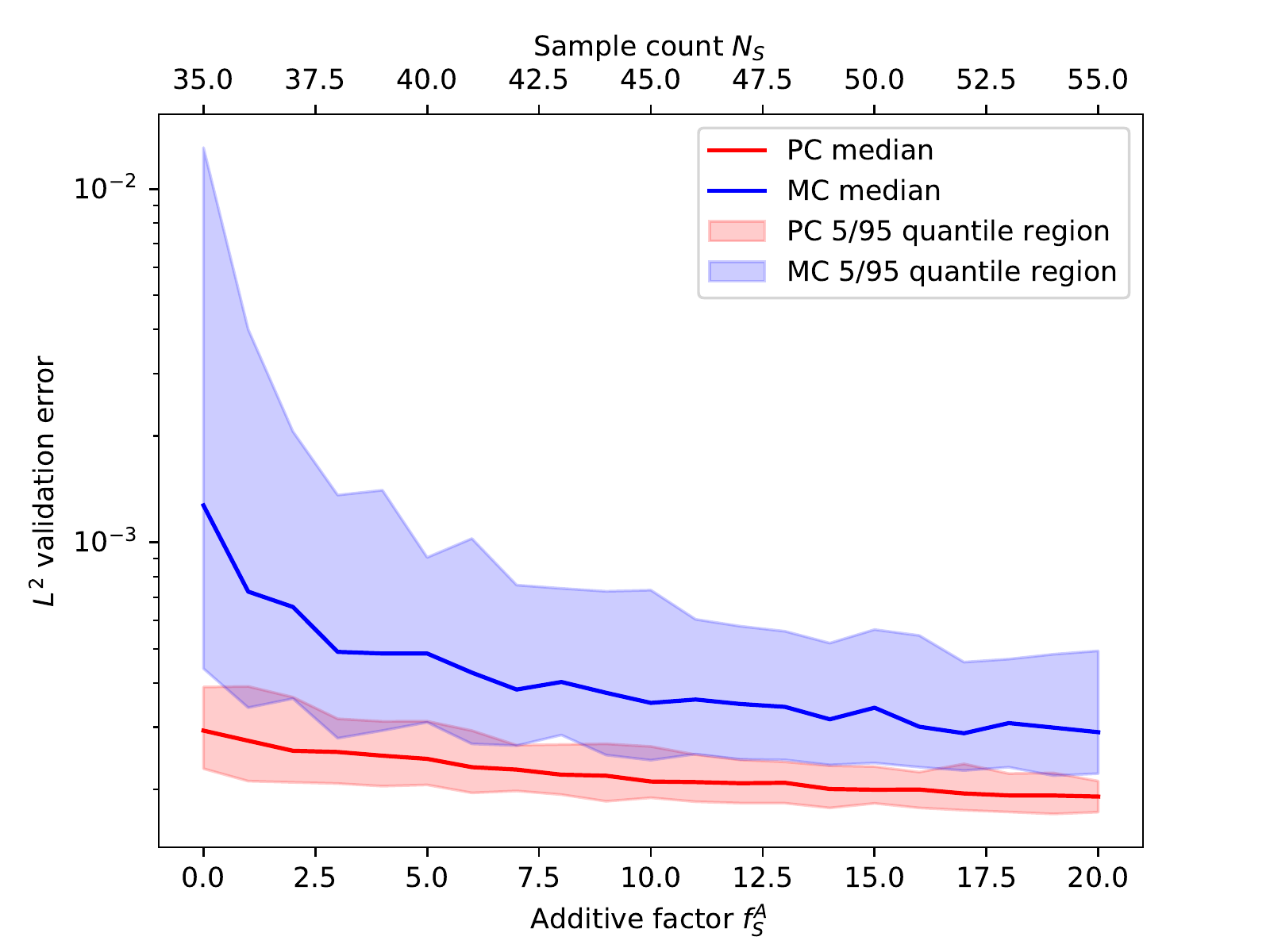}
\includegraphics[width=0.49\textwidth{}]{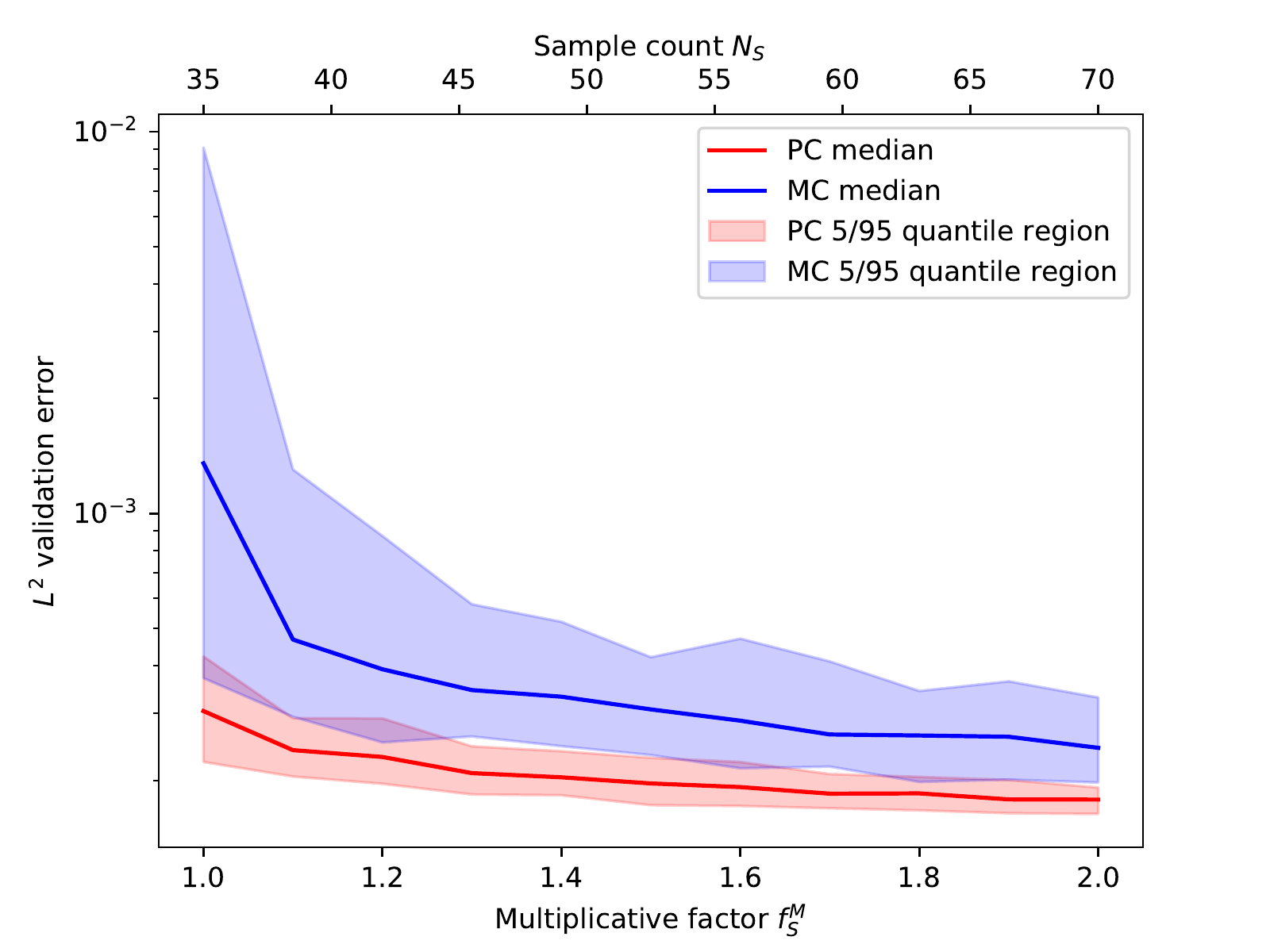}
\caption{
\revision{
  $L^2$ validation error on the \pfifty{} model for the WAFP PCE procedure proposed in this manuscript (``PC", red) versus a PC construction that uses iid Monte Carlo samples (``MC", blue). Left: $N_S = N_p + f_S^A$ samples. Right: $N_S = f_S^M N_p$ samples. The validation error is computed over $5000$ iid samples generated independently of the sampling procedures. Since the experimental design is random for both of these sampling procedures, quantile regions were computed using $100$ realizations of each kind of PCE construction. The PC procedure considered in this manuscript performs notably better than a standard MC approach.
}
}
\label{fig:l2error_AM}
\end{figure}

\revision{
First we assess both our choice to perform sampling via the WAFP approach, and our recommendation of choosing $N_S = N_p + f_S^A$ samples, with $f_S^A = 10$. Figure \ref{fig:l2error_AM} shows the behavior of WAFP sampling both as $f_S^A$ is varied for $N_S = N_p + f_S^A$ samples, and as $f_S^M$ is varied for $N_S = f_S^M N_p$ samples. The results demonstrate that the error depends mildly on $f_S^A$ once $f_S^A \gtrsim 5$. For safety, we therefore choose $f_S^A = 10$. If we instead consider a sample count $N_S = f_S^M N_p$, the results demonstrate that one can also attain attractive error results for say \revisionthree{$f_S^M \gtrsim 1.3$}, but this generally results in many more samples, so that the additive recommendation $N_S = N_p + 10$ achieves similar error with substantially reduced cost.

Figure \ref{fig:l2error_AM} also compares the WAFP strategy against error for an alternative PC construction built from $N_S$ iid Monte Carlo samples. The results show that the $WAFP$ strategy consistently outperforms the Monte Carlo strategy.
}

\begin{figure}[htbp]
\centering
\includegraphics[width=0.8\textwidth{}]{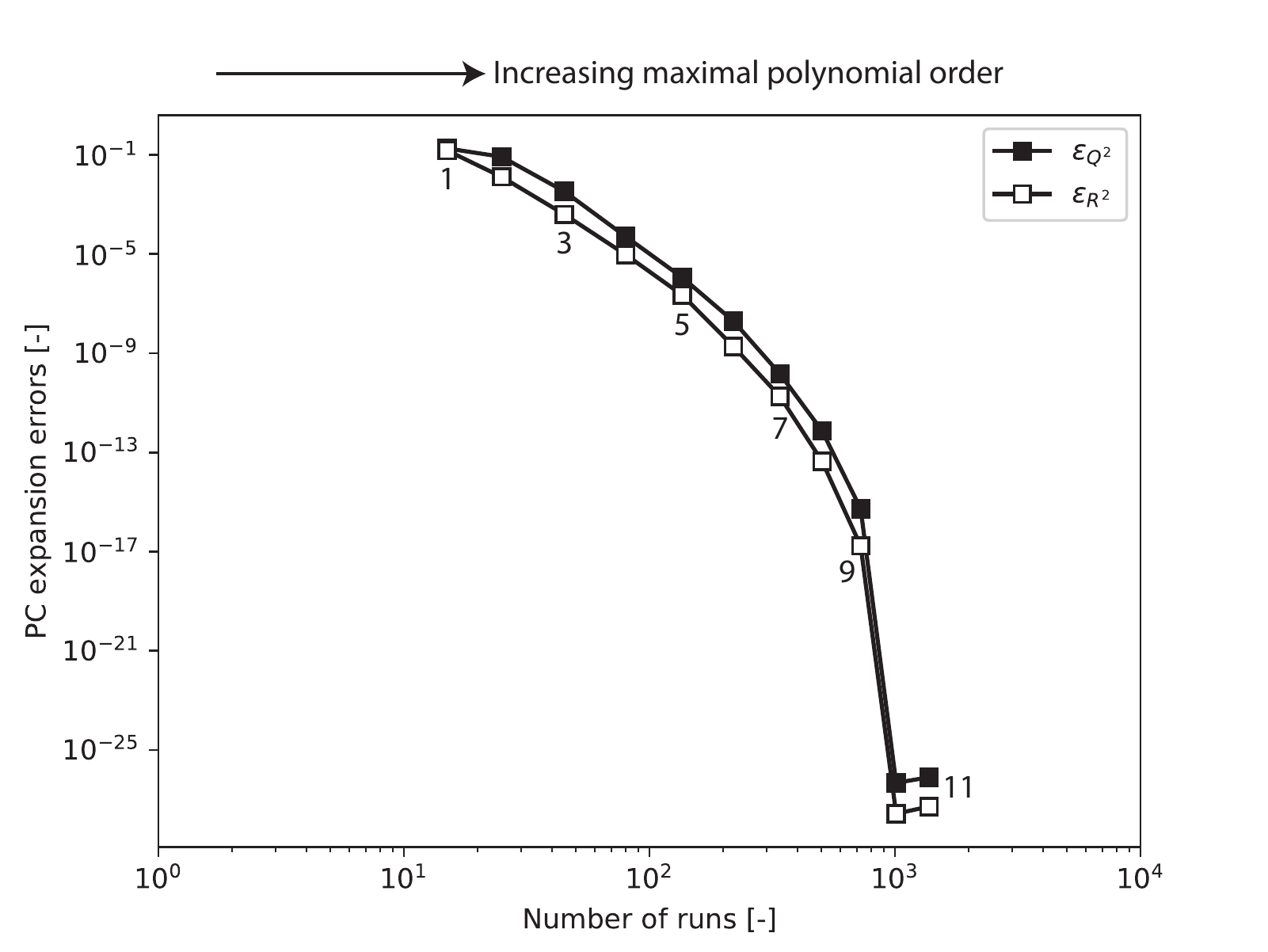}
\caption{Predictive ($\epsilon_{Q^2}$) and descriptive ($\epsilon_{R^2}$) errors for the \pce{} of the oxyhemoglobin dissociation model.}
\label{fig:pce_err}
\end{figure}

Predictive ($\epsilon_{Q^2}$) and descriptive errors ($\epsilon_{R^2}$) for the \pce{} are shown in Figure \ref{fig:pce_err}.  A maximum polynomial order of at least $p=3$ was required to obtain errors less than $10^{-2}$.  \pce s with order $p \geq 3$ had good predictive ($\epsilon_{Q^2}\ll 1$) sampling power for all q.  Discrepancy between the \pce{} description and the true model was low ($\epsilon_{R^2}\ll 1$) with maximal polynomial order $p \geq 3$.

\subsection{Sensitivity indices convergence and quality}

\begin{figure}[htbp]
\centering
\includegraphics[width=0.8\textwidth{}]{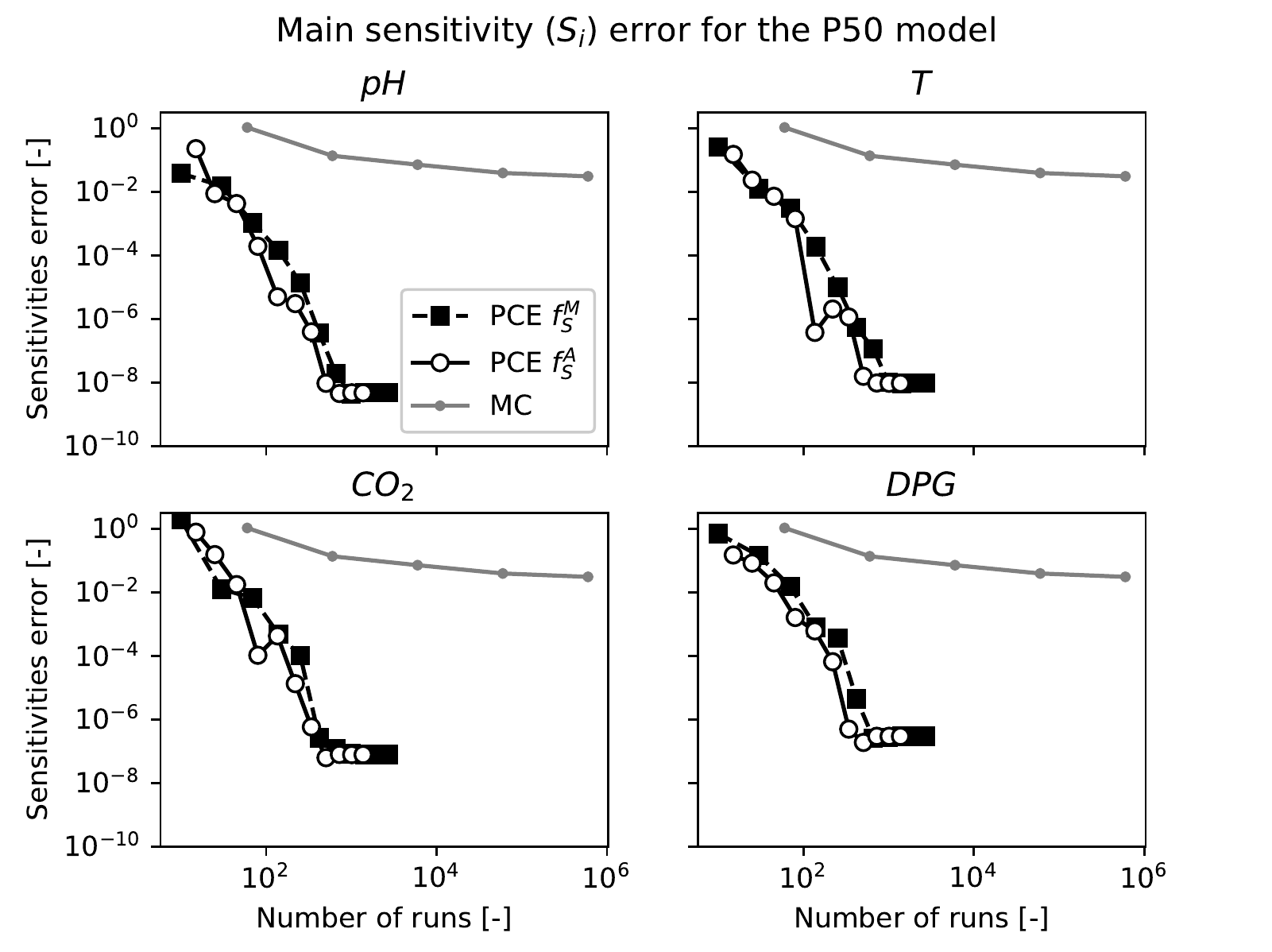}
\caption{Relative error ($\epsilon_{\delta{}}$) in the main sensitivity ($S_i$) indices for $P_{50}$ obtained using \pce s with additive and multiplicative sampling factors and with increasing polynomial orders.  The same relative error is plotted of the reference \mc{} analysis for $N=\{10,100,1000,10000,100000\}$ samples per parameter.}
\label{fig:ind_l2err_main}
\end{figure}

\begin{figure}[htbp]
\centering
\includegraphics[width=0.8\textwidth{}]{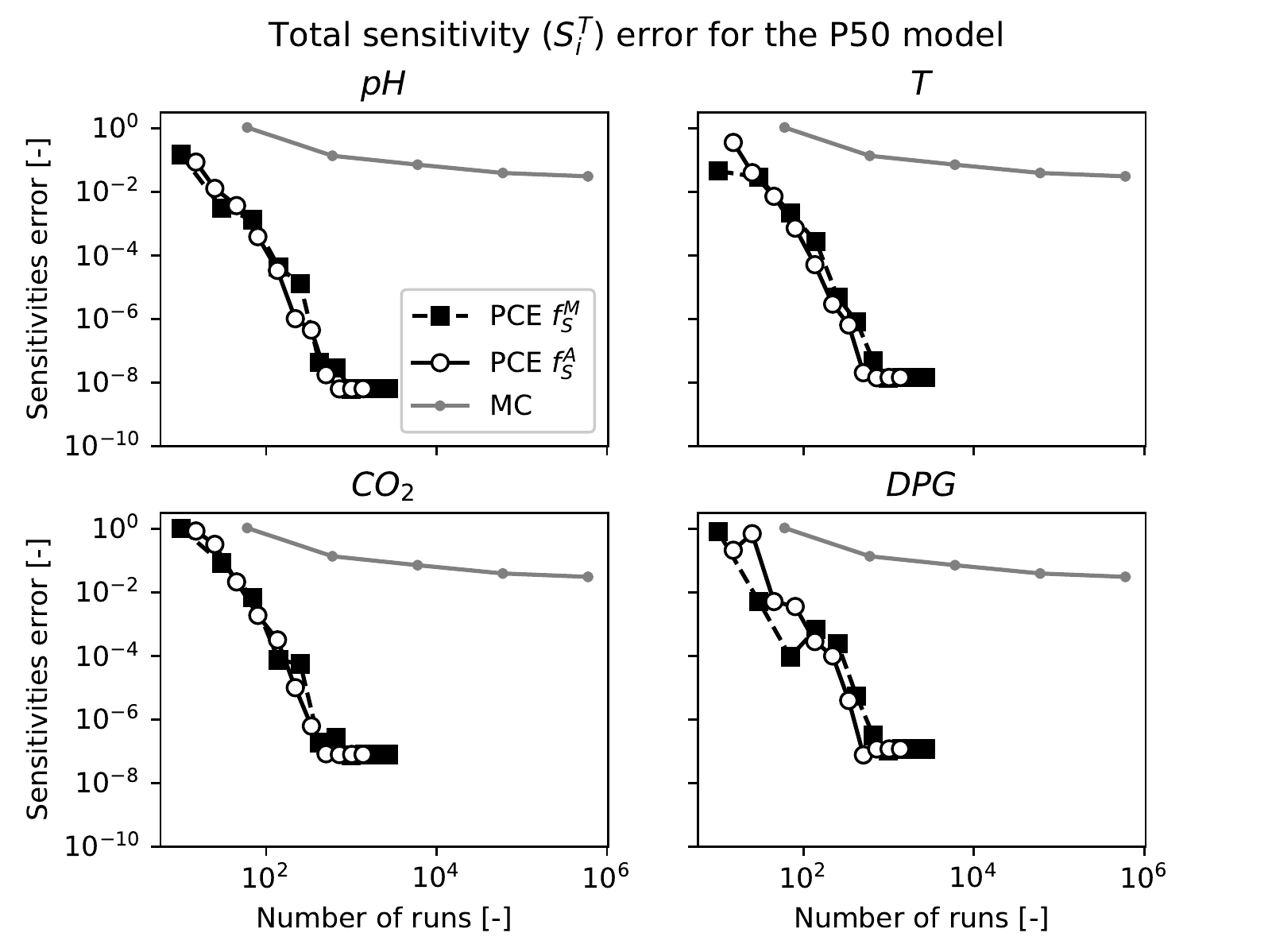}
\caption{Relative error ($\epsilon_{\delta{}}$) in the total sensitivity ($S_i^T$) indices for $P_{50}$ obtained using \pce s with additive and multiplicative sampling factors and with increasing polynomial orders.  The same relative error is plotted of the reference \mc{} analysis for $N=\{10,100,1000,10000,100000\}$ samples per parameter.}
\label{fig:ind_l2err_tot}
\end{figure}

Relative error ($\epsilon_{\delta{}}$) of the sensitivity indices for the additive and multiplicative \pce s are shown in Figures \ref{fig:ind_l2err_main} and \ref{fig:ind_l2err_tot}.  In general, both \pce s required a maximal polynomial orders $p=3$ to obtain a sensitivity error less than $10^{-2}$.  Although \mc{} analysis performed orders of magnitude more model runs, it did not obtain a sensitivity error less than $10^{-2}$.

\subsection{Oxyhemoglobin dissociation model}

\subsubsection{Uncertainty quantification}
\label{results:uq}

\begin{figure}[htbp]
\centering
\includegraphics[width=0.8\textwidth{}]{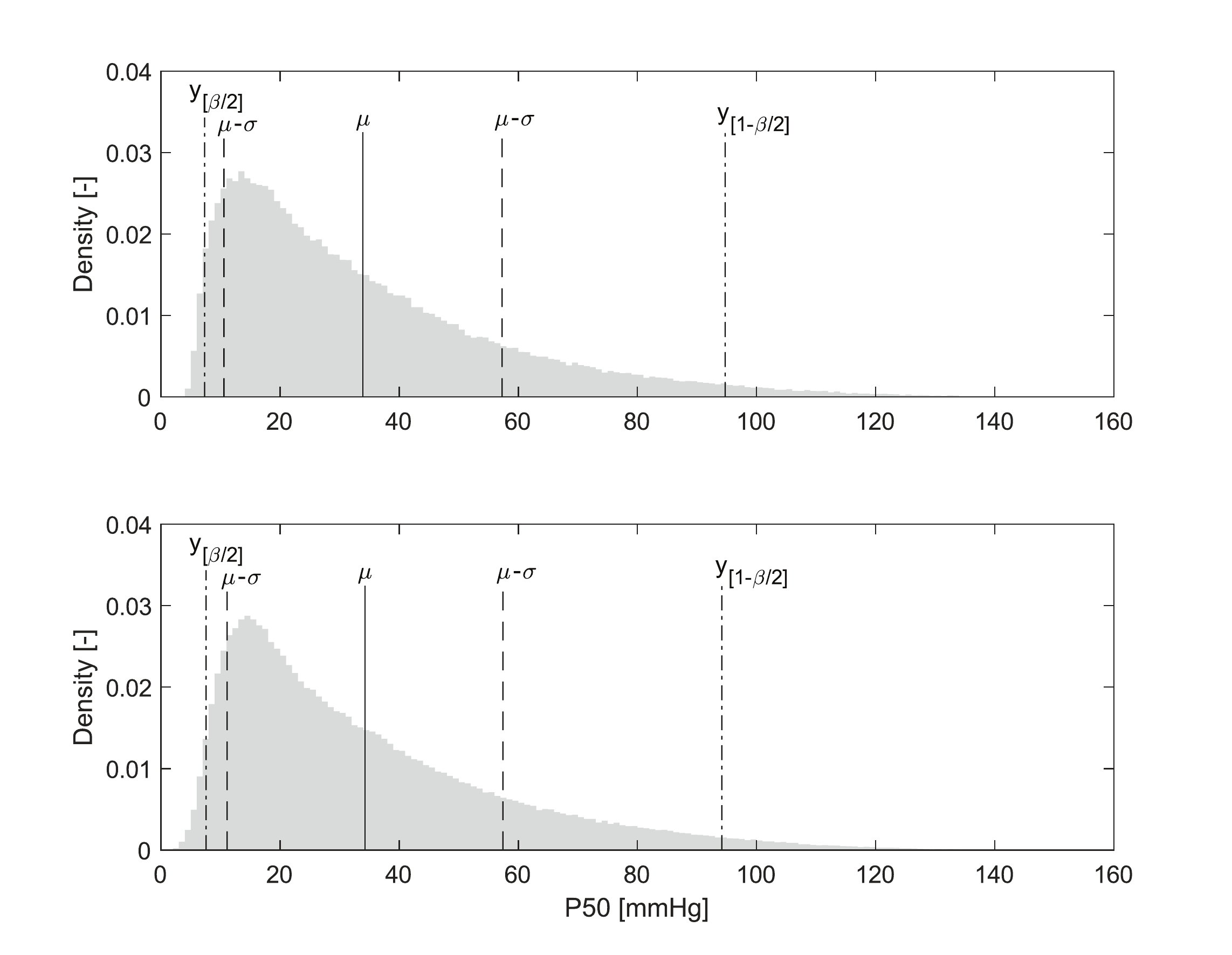}
\caption{Numerical evaluation of the distribution for \pfifty{} using Monte Carlo (\mc{}) and polynomial chaos (\pc{}) techniques.  \mc{} with $N_s=100,000$ is shown on top and \pc{} with order $p=3$ and $f_S^A=10$ is shown on bottom.  The center vertical thick line denotes the estimated value $\mu(P_{50})$.  The dashed lines to either side denote $\mu(P_{50})\pm{}\sigma(P_{50})$.  The dot-dashed lines represent the 95\% prediction interval.}
\label{fig:dist}
\end{figure}

The estimated distribution of \pfifty{} was similar for both methods (Figure \ref{fig:dist}).  The mean, standard deviation, and 95\% prediction intervals for \mc{} and \pc{} methods are shown in Table \ref{table:uq}.  When model inputs are sampled uniformly across a realistic physiological sample space, \pfifty{} does not have a normal distribution.  The distribution is instead asymmetric with positive skewness.

\subsubsection{Sensitivity analysis}
\label{results:sa}

\begin{figure}[htbp]
\centering
\includegraphics[width=0.7\textwidth{}]{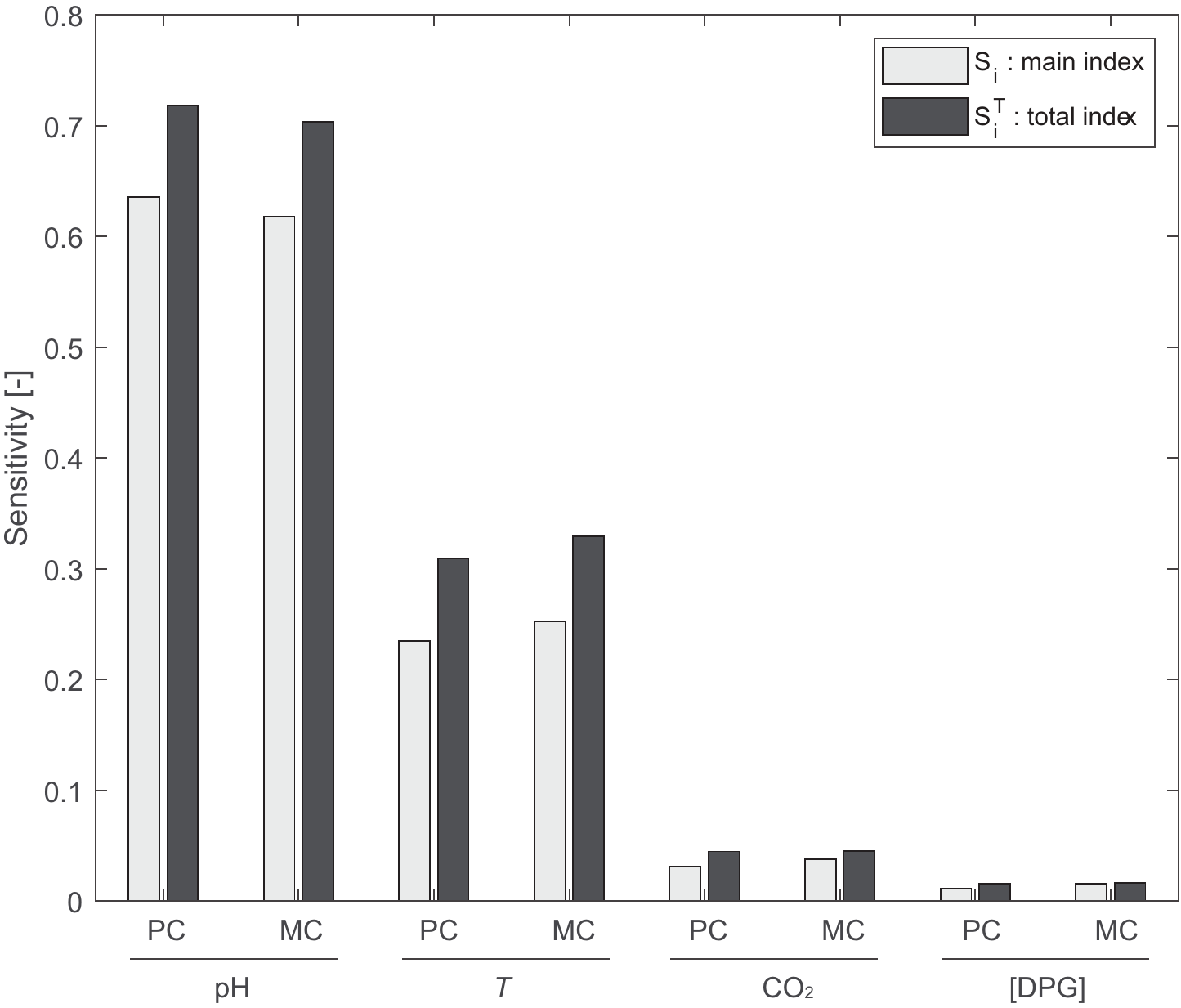}
\caption{Main ($S_i$) and total ($S^T_{i}$) sensitivity Sobol indices for pH, \tmp{}, \pcotwo{}, and \dpg{} with \pfifty{} as the output of interest.  Estimates from polynomial chaos (\pc{}) method are shown with the left two bars of each group.  Estimates from the Monte Carlo (\mc{}) method are shown with the right two bars.  For each method, $S_i$ is shown on the left and $S^T_{i}$ on the right.}
\label{fig:global_sens_analysis}
\end{figure}

The sensitivity indices calculated using \pc{} and \mc{} methods correspond well (Figure \ref{fig:global_sens_analysis}).  The parameter ranking for influence on model output was the same when using both methods.  For the concentration of hydrogen ions (\ph{}) and temperature (\tmp{}), both methods also reported total sensitivities that were different from main sensitivities, indicating a substantial interaction between the input variables.  

The concentration of hydrogen ions (\ph{}) was estimated as the most influential variable, followed by temperature (\tmp{}), the partial pressure of carbon dioxide (\pcotwo{}), and the concentration of \twothreedpg{} ([DPG]).  For these estimates, the concentration of hydrogen ions ($S^T \approx 0.72$) had more than twice the influence on total output variance than the second-ranked variable, temperature ($S^T \approx 0.31$).  The influence of the concentration of \twothreedpg{} on model output variance was calculated as negligible.

The sum of the main sensitivities for individual variables and interaction among pairs was $>0.99$, accounting for nearly all of the variance in the model's output.  Mutual interactions among paired variables accounted for $8.64\%$ of \revision{variance in \pfifty{}}, 7.8\% of which was attributed to pairwise interactions between \ph{}, \tmp{}, and \pcotwo{}.  One of the interactive effects had more influence on uncertainty of \pfifty{} than individual variables.  The interaction between the pair (\ph{}, \tmp{}) was ranked higher ($S \approx 0.07$) than both \pcotwo{} ($S \approx 0.03$) and \twothreedpg{} ($S \approx 0.01$).

\subsubsection{Global \uqsa{}}

\begin{figure}[htbp]
\centering
\includegraphics[width=0.7\textwidth{}]{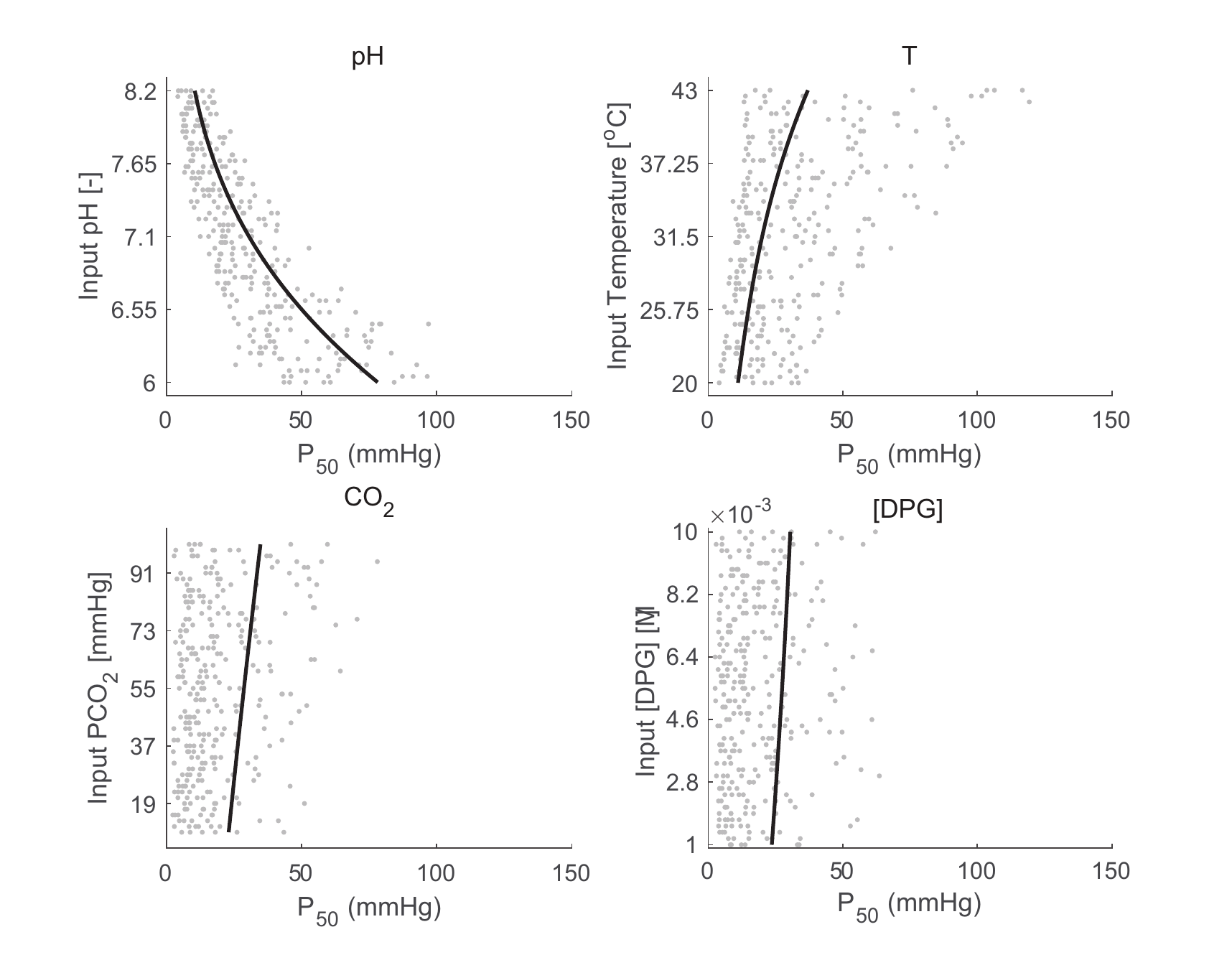}
\caption{Graphical portrayal of the non-linear output behavior for the oxyhemoglobin dissociation model which has four interacting model inputs, \ph{}, \tmp{}, \pcotwo{}, and \twothreedpg{}.  The solid black line indicates the value of each individual parameter across its physiological range with the three other parameters at standard values.  The gray dots represent the output $P_{50}=f(pH,T,PCO_2,[DPG])$, plotted as the individual parameter against the output, with the  values of the other parameters varying across their physiological range.  As shown, the response in-between each parameter's endpoints is non-linear, indicating that the use of a \pce{} would be advantageous as opposed to Monte Carlo methods}
\label{fig:odc_cumulative}
\end{figure}

Although for variables the extreme \pfifty{} values occurred at the ends of each range, the dependency of the equation of state output is non-linear \revisionthree{but smooth} across the entire input parameter space (Figure \ref{fig:odc_cumulative}).  \revisionthree{The smoothness of the map therefore makes} a \pce{}-based global \uqsa{} an appropriate analysis for this model.\cite{xiu2010numerical}  The \revisionthree{smooth} shape of the dependency justifies the use of a global \pce{} as opposed to \mc{} methods.  Since the relationship is \revisionthree{smooth}, \mc{} simulation techniques would not be as effective or accurate in providing a sample representative of the parameter space.  On the other hand, a \cfp{} based \pce{} efficiently provides an accurate emulator of \revisionthree{smooth} models.  


\section{Discussion}
\label{sec:discuss}

In this paper, we have proposed a procedure for computing the output uncertainty and sensitivity indices of physiological models with reduced computational cost.  This procedure reduces computational cost by constructing an inexpensive surrogate model using \pce{}.  This paper focuses on reducing the computational cost of constructing a \pce{} by proposing a deterministic least squares regression sampling procedure, Christoffel-based weighted approximate Fekete points (\cfp{}).  

The results show that using an additive sampling factor performs comparably to a multiplicative sampling factor.  For both additive and multiplicative methods, a maximal polynomial order $\geq 3$ was required to obtain sensitivity error lower than $10^{-2}$ \revision{indicating that the order of polynomial necessary was independent of sampling strategy.}.  However, an additive sampling factor is more efficient since it requires fewer model runs \revision{and can achieve the same accuracy using a fewer number of points}.  This becomes particularly important when $p \gg 1$ or $k \gg 1$, where $p$ is the \pce 's maximal degree and $k$ is the number of independent stochastic input parameters.

Implementing a \cfp{} based \pce{} can yield results that are the same as \mc{} methods.  However, the computational cost is much lower with \cfp{} based \pce{}.  One reason the computational cost is lower is because \cfp{} based \pce{} is able to use an additive sampling factor.  Using an additive sampling factor means that \cfp{} based \pce{} sampling size grows less as the maximal polynomial order and model dimensionality grow.

In this paper, we have used uncertainty quantification to analyze a four parameter model for describing changes in oxyhemoglobin dissociation as represented by \pfifty{}.  This uncertainty analysis has included analysis of the forward uncertainty propagation of \pfifty{} and of the sensitivity of \pfifty{} to input parameters \ph{}, \tmp{}, \pcotwo{}, and \twothreedpg{} as expected using the oxyhemoglobin dissociation model.  This quantification has shown that the expected distribution of \pfifty{} when using the model is positively skewed with the 95\% prediction interval for \pfifty{} ranging from 8 to 94 mm Hg. This analysis has shown that, as implemented in the model, \pfifty{} is most sensitive to perturbations of \ph{}, sensitive to \tmp{}, and insensitive to \pcotwo{} and \twothreedpg{}.  Therefore, only \ph{} and \tmp{} are interesting for modeling the uncertainty in \pfifty{} due to physiological variability.  \revision{Based on our simulation results using the \cfp{} procedure,} modeling \pcotwo{} and \twothreedpg{} does not contribute significantly to uncertainty in \pfifty{}.

The highest ranked parameter for sensitivity of Dash's equation of state aligns with theoretical and expert knowledge.  \ph's large influence on oxyhemoglobin dissociation, called the Bohr Effect, plays a major physiological role in transporting oxygen.  The Bohr Effect is the mutual interaction of oxygen binding and hydrogen ion binding of oxyhemoglobin \cite{siggaard1971oxygen}.  This mutual interaction allows a decrease in \ph{} near tissue capillaries to increase delivery of \otwo{} at the tissue level \cite{dash2016simple,rhoades2012medical}.  Thus the high influence of \ph{} on \pfifty{} plays a critical role in hemoglobin's ability to unload \otwo{} to tissues.

The pairwise interactions between influential parameters shown using sensitivity analysis should reflect theoretical and expert knowledge on how these parameters influence each other.  Dash's equation of state showed substantial interaction between \ph{} and \tmp{}.  This result aligns with experimental results from studies conducted separately from the studies used to establish the equation of state.  The interaction between these parameters has been described previously and shown to have a linear relationship \cite{rosenthal1948effect}.

Dash \textit{et al.} have developed an accurate equation of state for determining \shbotwo{} under various physiological conditions.\cite{dash2016simple} Dash's equation of state has been well-validated on a large set of experimental data from separate experiments available in the literature.\cite{joels1958carbon,naeraa1963influence,bauer1972carbamino,hlastala1977influence,matthew1977quantitative,reeves1980effect} Dash has used these data to characterize the dependence of \pfifty{} on \ph{}, \tmp{}, \pcotwo{}, and 2-3-diphosphoglycerate (\twothreedpg{}).  This equation of state is the most extensively validated using independent data and the most accurate at the time of this writing.  When implementing their equation of state, their model provided accuracy greater than 99.5\% over the whole saturation range when compared with experimental data.

Despite these efforts and this accuracy, the oxyhemoglobin dissociation model does not entirely align with theoretical and expert knowledge.  The main ($S \approx 0.01$) and total ($S^T \approx 0.02$) sensitivities for \twothreedpg{} do not support well-established knowledge that \twothreedpg{} alters \pfifty{} both by binding to hemoglobin and by altering pH \cite{thomas1974oxyhemoglobin,benesch1967effect,chanutin1967effect,tyuma1969different}.  Analysis of the oxyhemoglobin dissociation model shows that for the model the influence of \twothreedpg{} is negligible compared to the influence of \ph{}, \tmp{}, and \pcotwo{}.  \revisionthree{One explanation for the lack of sensitivity to \twothreedpg{} could be that variations in the normal physiological range do not change the output even though the unphysiological range close to 0 would drastically affect the oxygen release from hemoglobin.} However, without \twothreedpg{} almost no oxygen would be released from hemoglobin.  This implies that \twothreedpg{}, either directly or indirectly, does in fact influence oxyhemoglobin dissociation.

Despite conservatively testing a broad range of \pcotwo{}, Dash's equation of state shows minimal interaction between \ph{} and \pcotwo{} ($S_{pH,CO_2} \approx 0.01$).  The interaction between \ph{} and \pcotwo{} in human blood is well-established and described by the Henderson-Hasselbach equation \cite{johnson2003essential}:

\begin{align}
	pH=pK_{a,H_2CO_3}+\textrm{log}_{10}\Big(\frac{[HCO_3^-]}{[H_2CO_3]}\Big).
\end{align}

\noindent{}This interaction is an essential part of the bicarbonate buffer system for maintaining \ph{} in the blood.  Without it the body would be unable to maintain acid-base balance.

Up to this point, the focus when developing oxyhemoglobin saturation models has been on the accuracy of the model across the full range of \potwo{}.  The analysis in this paper shows the importance of also performing \uqsa{} to analyze the fidelity of a model.  Despite high accuracy when analyzed using experimental data, the Dash model only partially aligns with clinical observation and expert knowledge on changes in oxyhemoglobin saturation.  Performing \uqsa{} has highlighted the characteristics of the model which require adjustment before model fidelity can be obtained.  Two of these areas include the influence on \pfifty{} of \dpg{} individually and of the interaction between \ph{} and \cotwo{}.  Only after model characteristics are improved in these two areas will the Dash model be able to properly simulate changes in \pfifty{} due to different levels of these effectors.

In their paper, Dash and Bassingthwaighte initially concluded that \ph{} and \tmp{} have the most influence on \otwo{} binding.\cite{dash2010erratum} After refining their model they concluded that \ph{} significantly influences \otwo{} binding compared to the smaller effects of \pcotwo{}, \twothreedpg{}, and \tmp{}.\cite{dash2016simple} \uqsa{} was not incorporated into Dash et al.'s studies and thus they did not reach their conclusions using formal quantitative computational methods nor did they provide a description of the interactions among variables.  We have shown the value of \uqsa{} by providing a precise quantitative description of the uncertainty of \pfifty{} and a ranking of the sensitivity of \pfifty{} to all four variables.  We have also shown the value of \uqsa{} by providing a formal analysis of the interactions among variables.

The Dash model is interesting to analyze because it has two influential (\ph{}, \tmp{}) and two noninfluential parameters (\pcotwo{}, \dpg{}).  It is also interesting because two of the variables interact.  These interesting aspects make the Dash model a good candidate for future benchmarking of uncertainty quantification techniques.

Many other models exist which describe oxyhemoglobin dissociation.  In future research, Buerk and Bridges' model \cite{buerk1986simplified}, which Dash and Bassingthwaighte based their initial model on, could be evaluated for comparison.  Further, the methods described in this paper could also be performed on models which incorporate different subsets of the four model input parameters analyzed here.  Analysis of these models could then be compared with the analysis in this paper to determine which models align best with real world observation.

We have assumed a uniform distribution of \ph{}, \tmp{}, \pcotwo{}, and \twothreedpg{} across the parameter space.  Right now there are no results providing evidence of a different distribution.  With a uniform distribution, any value of \ph{}, \tmp{}, \pcotwo{}, and \twothreedpg{} is equally likely to be sampled.  Should data describing the distribution of one or all of these parameters become available in the future, that distribution could be used in a \pce{}, and an analysis identical to the one presented here could be used to study uncertainty and sensitivity, \revision{with appropriate modifications to the input distribution}.

The focus of this paper has been on implementing a \cfp{} based \pce{} procedure for reducing computational cost.  Other methods for reducing computational cost have been described previously including input parameter screening prior to constructing a \pce{} as well as sparse \pce{}.\cite{donders2015personalization,blatman2010adaptive}  Although this paper has focused on \cfp{} based \pce{}, ideally a combination of these methods could be used simultaneously to further reduce the computational cost of \pce{}.

\section{Conclusion}\label{sec:cnclsn}

We have proposed a procedure for constructing a \pce{} using weighted approximate Fekete points and analyzed a model describing oxyhemoglobin dissociation using this procedure.  The procedure reduces the required sampling factor for performing \pce{} from multiplicative to additive.  Results from the particular physiological model used in this paper have shown that, when based on weighted approximate Fekete points, a \pce{} using an additive sampling factor performs comparably to one using a multiplicative factor. Such a result suggests that a moderate amount of data is enough to construct a meaningful \pce{} for prediction and UQ analysis.  This procedure is also useful for comparing existing models to clinical observation and expert behavior in order to analyze model fidelity.


\setcounter{table}{0}
\renewcommand{\thesubsection}{A.\arabic{subsection}}

\section*{Appendix A: Benchmark Testing}
\label{sec:appndx}

There are several test functions which are useful for validating a particular implementation of \pce{} because their sensitivity values can be derived analytically.\cite{sudret2008global}  The functions considered here are a polynomial model, the Ishigami function, and the G-function.  Since these functions have analytical sensitivity values, they are useful for benchmark testing implementation of \pce{} computer methods.  Sudret \etal{} have reported the analytical sensitivity values for these functions previously.\cite{sudret2008global} Here we benchmark test a \pce{} as proposed in this paper to confirm the \pce{} performs adequately.  Relative error for estimates of the sensitivity indices was calculated using (\ref{eq:relerr}).

\subsection{Polynomial model}

First, we consider the model:

\begin{align}\label{eq:polymdl}
Y=\frac{1}{2^n}\prod_{i=1}^{n}(3X_i^2+1), \tag{A.1}
\end{align}

\noindent{}for n = 3 where the input variables $X_i$ are uniformly distributed over [0, 1].  A polynomial model is useful for benchmark testing to determine whether the \pce{} procedure can produce exact results.  Given the model in (\ref{eq:polymdl}) is a polynomial of degree 6, a maximal order of $p=6$ is chosen.  When using the procedure outlined in this paper, including an additive sampling factor $f_S^A=10$, the exact values for the sensitivities were found (tested to 12-digit accuracy).

\revision{One challenge with PC methods is the choice of polynomial order $p$. Non-adaptive and adaptive methods are common; in this manuscript we choose non-adaptive methods for simplicity, where $p$ is chosen \textit{a priori}. Adaptive methods that refine the polynomial approximation space are more flexible since in principle they can be used to drive PC error down to a desired tolerance; however, both computational strategies and convergence for adaptive methods is still an active area of research.}
To gain an understanding of how the choice of polynomial order $p$ can affect simulation accuracy, we test expansions with $p=3,4,5$ (see Table \ref{tab:polymdl}).  A \pce{} with maximal order $p=3$ is able to estimate the main sensitivities with a relative error less than 1\% while the second-order and total sensitivity relative errors were within 5\% (absolute error within $\pm 0.01$).  A \pce{} with maximal order $p=4$ estimated the main and second-order sensitivities with a relative error within 1\% and the largest total sensitivity error of 9.00\% (absolute error within $\pm 0.001$).  With maximal order $p=5$, the \pce{} estimated all sensitivities with relative error within 0.5\%.

\subsection{Ishigami function}

Next, we consider the Ishigami function:

\begin{align}
	Y=\textrm{sin}X_1+a\textrm{sin}^2X_2+bX_{3}^4\textrm{sin}X_1, \tag{A.2}
\end{align}

\noindent{}where input variables $X_i$ are uniformly distributed over [-$\pi$, $\pi$] with $a=7$ and $b=0.1$.  All three input variables of the Ishigami function are influential (whether directly or through interactions).  Although this is uncommon in biomedical engineering applications, this characteristic is useful for benchmark testing because it typically requires a high order of maximal polynomial degree.  The Ishigami function is also commonly used for benchmark testing because of its strong nonlinearity and nonmonoticity.

For this function, we tested expansions with maximal order $p=\{3, 5, 7, 9\}$ (see Table \ref{tab:ish}). With a threshold of 0.06 for identifying influential sensitivities, a maximal order $p=5$ is required to identify influential parameters and parameter interactions correctly (maximum relative error of 40\%).  A maximal order $p=7$ obtains all sensitivities with a relative error within 10\%.  Increasing the maximal order to $p=9$ estimates all sensitivities with a relative error within 1\%.

\subsection{G-function}

Lastly, we consider Sobol's G-function:

\begin{align}
Y=\prod_{i=1}^{q}\frac{|4X_i-2|+a_i}{1+a_i}, \tag{A.3}
\end{align}

\noindent{}where input variables $X_i$ are uniformly distributed over [0, 1], $q=8$, and $\mathbf{a}=\{1, 2, 5, 10, 20, 50, 100, 500\}$.  With $q=8$, this model can be used to first screen for influential parameters using a low maximal order \pce{} after which the dimension of the model can be reduced by fixing noninfluential parameters to their mean value (0.5 in this case).  Reducing the dimension lowers the computational cost for constructing a higher maximal order \pce{} focused on estimating sensitivities for influential parameters only.

For the G-function, we screened for influential parameters using a maximal order $p=2$ which required 55 model runs.  Threshold values of 0.02 for main sensitivity and 0.07 for total sensitivity were required to correctly distinguish between influential ($X_1-X_4$) and noninfluential ($X_5-X_8$) parameters (see Table \ref{tab:sobolfull}).  As planned, This benchmark test then proceeded by setting $X_5-X_8$ to their mean value of 0.5, effectively reducing the dimensions of the model, after which a \pce{} representing the reduced model was constructed to further investigate the sensitivity of the model output to input parameters $X_1-X_4$.

For the reduced G-function, we tested expansions with maximal order $p=\{3, 5, 7, 9\}$ (Table \ref{tab:sobolred}).  A maximal order of $p=5$ is required to rank influential sensitivities correctly. Increasing the maximal order to $p=9$ estimated all main and second-order sensitivities with an absolute error within 0.02 and total sensitivities with an absolute error within 0.04. 

\subsection{Conclusion}

We have used benchmark testing to analyze the quality of the procedure for constructing a \pce{} as described in this paper.  This \pce{}, which uses an additive sampling factor, performed comparably to other \pce s previously tested and reported in the literature.  Therefore, we conclude that it is prepared for performing analysis of additional models including the model for describing oxyhemoglobin dissociation.

\section{Acknowledgements}

\bibliographystyle{siam}
\bibliography{myrefs}

\end{document}